\documentclass[aps,prd,amsmath,floats,floatfix, twocolumn,
superscriptaddress,nofootinbib,showpacs,longbibliography]{revtex4-1}
\usepackage[T1]{fontenc}
\usepackage[utf8]{inputenc}
\usepackage{lmodern}
\usepackage{verbatim}
\usepackage[dvipsnames, usenames]{xcolor}
\definecolor{linkcolor}{rgb}{0.0,0.3,0.5}
\usepackage[hypertexnames=false, unicode, colorlinks=true, linkcolor=linkcolor,
citecolor=linkcolor, filecolor=linkcolor,urlcolor=linkcolor,
pdfusetitle]{hyperref}
\usepackage{amsmath,amssymb,graphicx,gensymb}
\usepackage{subfigure}
\usepackage[normalem]{ulem}
\usepackage{mathtools}

\definecolor{orange}{rgb}{1,0.5,0}
\definecolor{mygray}{gray}{0.9}
\definecolor{darkgreen}{rgb}{0.33, 0.42, 0.18}

\def\btheta{{\bm \theta}}
\def\PL{\phi_\mathrm{lens}}
\def\DT{\Delta t_{12}}
\def\MG{\text{LIB}}
\def\GR{\text{GR}}
\def\BF{\ensuremath{\mathcal{B}^{\MG}_{\GR}}}
\begin{document}

\title{Probing lens-induced gravitational-wave birefringence as a test of general relativity }

\author{Srashti Goyal}
\email{srashti.goyal@icts.res.in}
\affiliation{International Centre for Theoretical Science, Tata Institute of Fundamental Research, Bangalore - 560089, India}

\author{Aditya Vijaykumar}
\email{aditya.vijaykumar@icts.res.in}
\affiliation{International Centre for Theoretical Science, Tata Institute of Fundamental Research, Bangalore - 560089, India}
\affiliation{Department of Physics, The University of Chicago, 5640 South Ellis Avenue, Chicago, Illinois 60637, USA}

\author{Jose Mar\'ia Ezquiaga}
\email{jose.ezquiaga@nbi.ku.dk}
\affiliation{Niels Bohr International Academy, Niels Bohr Institute, Blegdamsvej 17, DK-2100 Copenhagen, Denmark}

\author{Miguel Zumalac\'arregui}
\email{miguel.zumalacarregui@aei.mpg.de}
\affiliation{Max Planck Institute for Gravitational Physics (Albert Einstein Institute) \\
Am Mühlenberg 1, D-14476 Potsdam-Golm, Germany}


\date{\today}

\begin{abstract}
Theories beyond general relativity (GR) modify the propagation of gravitational waves (GWs). 
In some, inhomogeneities (aka. gravitational lenses) allow interactions between the metric and additional fields to cause lens-induced birefringence (LIB): a different speed of the two linear GW polarisations ($+$ and $\times$). Inhomogeneities then act  as  non-isotropic crystals, splitting the GW signal into two components whose relative time delay depends on the theory and lens parameters. Here we study the observational prospects for GW scrambling, i.e when the time delay between both GW polarisations is smaller than the signal's duration and the waveform recorded by a detector is distorted. 
We  analyze the latest LIGO--Virgo--KAGRA catalog, GWTC-3, and find no conclusive evidence for LIB. 
The highest log Bayes factor that we find in favour of LIB is $3.21$ for GW$190521$, a particularly loud but short event. However, when accounting for false alarms due to (Gaussian) noise fluctuations, this evidence is below 1--$\sigma$.
The tightest constraint on the time delay is $<0.51$ ms (90\% C.L.) from GW$200311\_115853$.
From the non-observation of GW scrambling, we constrain the optical depth for LIB,
accounting for the chance of randomly distributed lenses (eg. galaxies) 
along the line of sight. Our LIB constraints on a (quartic) scalar-tensor Horndeski theory are more stringent than solar system tests for a wide parameter range and comparable to GW170817 in some limits. Interpreting GW190521 as an AGN binary (i.e. taking an AGN flare as a counterpart) allows even more stringent constraints.  Our results demonstrate the potential and high sensitivity achievable by tests of GR, based on GW lensing.
\end{abstract}

\maketitle

\section{Introduction}\label{sec:introduction}

The detection of gravitational waves (GW) 
using the LIGO--Virgo--KAGRA (LVK) detectors \cite{LIGODet,VirgoDet,KAGRADet} 
from mergers of compact objects~\cite{gw150914,gw170817,gwtc1,LIGOScientific:2020ibl, LIGOScientific:2021djp,Nitz:2019hdf, Nitz:2021zwj, Venumadhav:2019lyq} has enabled precision tests of general relativity (\GR) in the strong field regime~\cite{LSC_2016grtests,GWTC1TestofGR,LIGOScientific:2020tif, LIGOScientific:2021sio}. 
Far away from the source, $\GR$ predicts that GWs are well described as linear perturbations of  the background Friedmann-Robertson-Walker (FRW) metric ~\cite{Will:1997bb} 
Existing propagation tests hence typically consider modifications over the FRW background and its effect on the GW signals as measured at the detectors \cite{Ezquiaga:2018btd, PhysRevD.97.104038, Nishizawa_2018}.

GR also dictates that GWs have only two tensor  polarisations ($+,\times$) which propagate independently from each other at the speed of light. 
However, in alternative theories of gravity, extra degrees of freedom (tensor, vector, scalar)~\cite{Will:2014kxa} can mix with GWs as they propagate, producing phenomena similar to neutrino oscillations (i.e. due to interactions between different neutrino flavors \cite{Workman:2022ynf}). In Lorentz invariant theories, the symmetries of the FRW restricts mixing effects to tensor degrees of freedom, either fundamental (e.g. in bigravity) or composite (e.g. multiple vector fields)~\cite{Max:2017flc,Max:2017kdc,Belgacem:2019pkk,Jimenez:2019lrk,Ezquiaga:2021ler}. However, inhomogeneities spontaneously break Lorentz symmetry, allowing interaction between GWs and scalar or vector degrees of freedom~\cite{Ezquiaga:2020dao,Dalang:2020eaj}.
This leads to new, testable predictions, and opens new opportunities to probe the gravitational sector beyond the FRW limit.

The evolution of GWs on an inhomogeneous background is described via propagation eigenstates: linear combinations of the interaction eigenstates ($h_+,h_\times$ and perturbations of additional polarisations) with a well-defined dispersion relation (analogous to massive neutrinos). 
As the relation between interaction and propagation eigenstates and their speed depends on position and direction, an inhomogeneous region of space splits the original signal into several components, each arriving with a relative time delay~\cite{Ezquiaga:2020dao}. Moreover, if deviations from $\GR$ are small, two eigenstates correspond to \textit{mostly-tensorial polarizations (linear combinations of $h_+,h_\times$ plus a negligible correction distinguishing both), with a very small speed difference.%
\footnote{We will ignore the remaining eigenstates (perturbations of beyond-$\GR$ fields plus negligible corrections) because 1) their emission needs to be suppressed to avoid dipolar radiation and 2) their speed can be substantially different, making an association with the the mostly-tensorial part of the signal difficult \cite{Ezquiaga:2020dao}. \label{foot:ignore_extra_pol}}
}

We will refer to the difference in propagation speed between the $+,\times$ polarizations as \textit{lens-induced birefringence} ($\MG$).
$\MG$ is analogous to the way a non-isotropic crystal, such as calcite, splits light in two beams. This splitting is caused by a difference in the refractive index of the linear electromagnetic polarizations, which depends on the alignment of the polarization vector with the crystal structure. In our case birefringence is caused not by anisotropies in a crystal, but by the background configuration of additional, non-$\GR$ fields which spontaneously break Lorentz symmetry. Moreover, $\MG$ splitting is independent of the frequency (in the high frequency approximation assumed), which would correspond to a perfectly isochromatic birefringent crystal. Because GW detectors have excellent time resolution and bad sky localization, our main observable will be the time delay between splitted signals and not their angular separation.

If the arrival time difference between the mostly-tensorial polarisations is larger than the duration of the binary merger signal then we would see only one polarisation at a time, mimicking a binary with either a zero inclination angle (face-on binary) or  with $90\degree$ inclination angle (face-off binary), appearing as GW echoes. 
Since the detectors are more sensitive to face-on binaries, one can expect excess of near zero inclinations in case of birefringence for the population of binaries. 
If the delay between the polarisations is larger than typical observing runs or the amplitude of one of the polarisations decays faster than the other, e.g. in Chern-Simons gravity \cite{Okounkova:2021xjv}, one would also expect an anisotropic inclination distribution. 
Current observations though show that the orientation distribution is consistent with 
being isotropic~\cite{Vitale:2022pmu}.  

However, when the time delay is less than the duration the signal, the GW waveform would be distorted or ``scrambled'' due to the interference of both polarisations.
Note that this effect is frequency-independent, and hence distinguishable from a different dispersion relation for the $+$ and $\times$ modes or the circularly polarized combinations (L-R), as predicted in \GR~\cite{deRham:2019ctd,Andersson:2020gsj,Oancea:2022szu} and alternative theories~\cite{Wang:2021gqm,Haegel:2022ymk,Bombacigno:2022naf}. As it is not suppressed by the frequency, $\MG$ is the dominant effect in the high-frequency limit for theories in which this effect is present.

Our study analyzes for the first time 
arrival time difference ($\DT$) between the two polarisation states due to different propagation speeds (frequency-independent
dispersion relations) as a result of $\MG$. This is a new, model-independent test of a basic prediction of $\GR$. We use these generic results to constrain GW lensing effects beyond \GR, for example in scalar-tensor theories with derivative interactions ~\cite{Ezquiaga:2020dao}. 

$\MG$ signatures are not linked to a specific regime of gravitational lensing in GR, such as strongly magnified or multiple images. The scale on which $\MG$ can be observed is very sensitive to the theory parameters and independent of the Einstein radius $R_E$, which characterizes the regimes of gravitational lensing. Hence, for sufficiently strong deviations from $\GR$, $\MG$ can be detected for impact parameters much larger than $R_E$, typically associated to weak lensing. Therefore, $\MG$-tests can be applied to all the GW detections. 
In addition, $\MG$ can be important for lenses very close to the source or the observer, for which $R_E$ vanishes. This is particularly interesting for sources merging near massive objects (e.g. a supermassive BH) since the background configuration of the additional fields enhances $\MG$.

The rest of the paper is organised as follows. In Sec.~\ref{sec:method} we describe our $\MG$ waveform model, methods for data analysis and introduce parameterized $\MG$ observation probabilities. In Sec.~\ref{sec:results}, we perform the birefringence test over a set of simulated GW events, and then to real events using the Bayesian model selection framework. In Sec.~\ref{sec: implications}, we study the implications of the results in constraining LIB probabilities and beyond-GR theories. Finally, in Sec.~\ref{sec:conclusion} we summarize the main results and discuss future prospects.

\begin{figure*}[tbh!]
    \centering
    \includegraphics[width=0.96\textwidth]{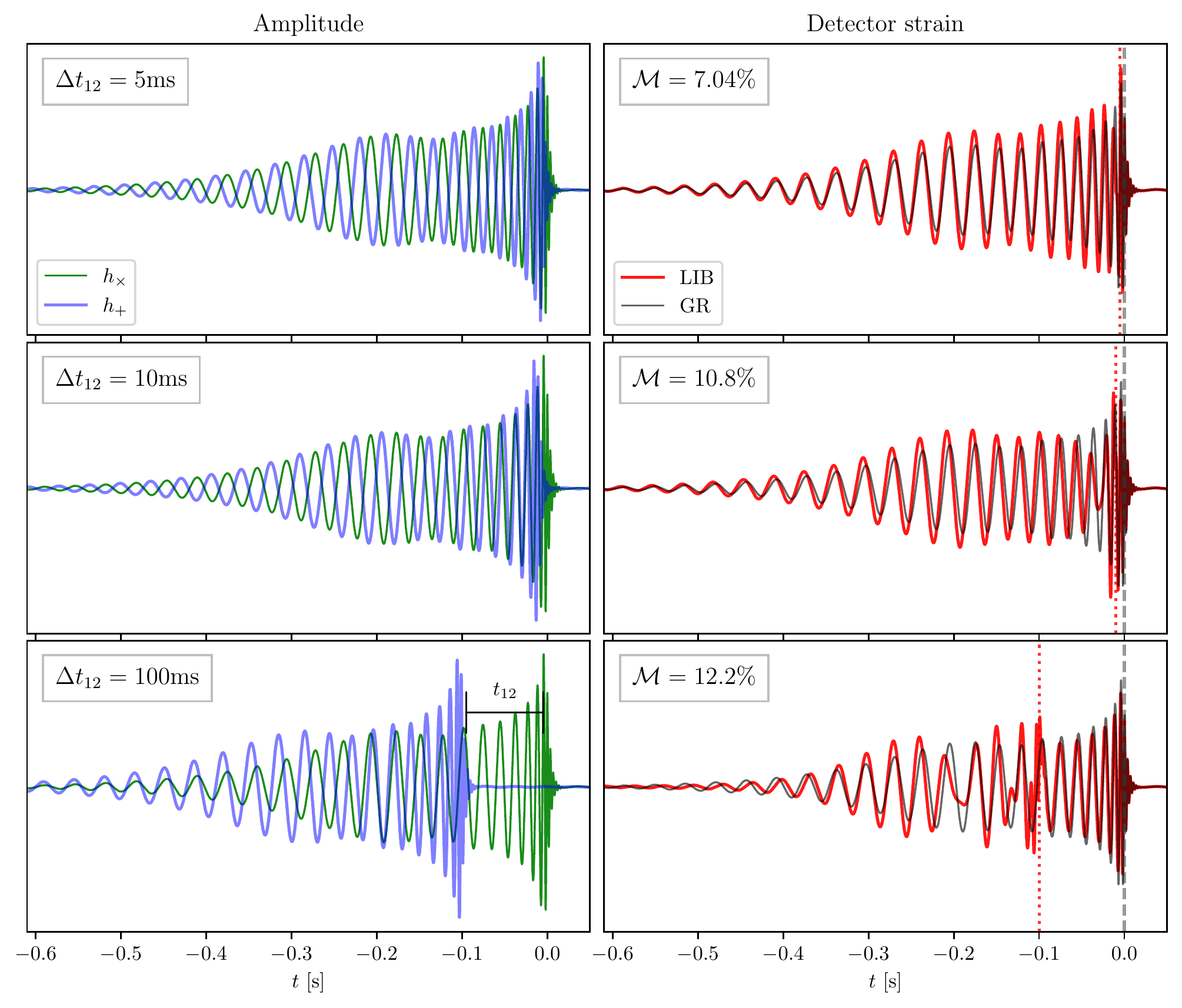}
    \caption{GW polarisations (left) and detector strain (right) for a CBC $(30+30)M_\odot$ with birefringent time delays $\Delta t_{12}=5, 10, 100$ ms (top to bottom). The sky localization and detector orientation correspond to $F^+= -0.38$, $F^\times= 0.71$ and $\MG$ strain is given by Eq. ~\ref{eq:hmodGRsimp} } 
    \label{fig:mg_waveform}
\end{figure*}

\section{Method}\label{sec:method}
In \GR, GWs have only two  polarisations $(+, \times)$ which propagate independently at the speed of light over the background FRW metric.
A given ground-based detector $I$ measures the GW signal, $h(t)$ as a linear combination of these polarisations~\cite{Veitch:2014wba},
\begin{equation}
    h_I(t) = F^{+}_I h_{+}(t) + F^{\times}_I h_{\times}(t)
    \label{eq:ht}
\end{equation}
where, $F^{+}_I, F^{\times}_I$ are the detector antenna pattern functions. 
In the case of compact binary coalescence (CBC), the relative amplitude of the polarisation modes depends on the inclination and polarisation angles of the binary  w.r.t to the line of sight, and also depends on the sensitivity of the detector for the source location at the time of arrival. The overall amplitude of the signal is inversely proportional to the luminosity distance of the source. Masses and spins of the source dictate the frequency evolution of the signal and its amplitude.

\subsection{Parameterized Lens-induced Birefringence Waveforms}\label{sec:phenom_model}

When there is any inhomogeneity along the travel path of a GW, e.g. an intervening galaxy, the GW can be gravitationally lensed. 
Gravitational lensing of a GW can produce multiple images of the original signal (strong lensing) or cause distortions (microlensing), but, in \GR, both polarisations are affected in the same way, i.e. the polarisation rotation is negligible for any sensible astrophysical lens \cite{Ezquiaga:2020gdt,Cusin:2019rmt}. 
However, in alternative theories of gravity the additional fields can couple with the tensor polarizations around  the lens and modify the GW propagation eigenstates. These eigenstates are a linear combination of original GW polarisations that evolve independently, each with a different speed, thus reaching the detectors at different times. 
We will assume spherically symmetric lenses, focus on the limit of small deviations from $\GR$, so the mostly-metric propagation eigenstates correspond to linear combinations of $h_+,h_\times$ (depending on the projected angle between the lens and the source), and neglect the additonal eigenstates (See Ref. \cite{Ezquiaga:2020dao} and footnote \ref{foot:ignore_extra_pol}).

This class of $\MG$ of GWs 
can be captured in a phenomenological manner as proposed in Ref.~\cite{Ezquiaga:2020dao}. 
After diagonalizing the propagation equations, the propagation eigenstates can be computed and one gets the transformation matrix 
${\mathcal S}$ relating the polarisation amplitudes in $\GR$ and after the $\MG$:
\begin{equation}
[h_{+},h_{\times}]^T_{\MG} = \mathcal{S} [h_{+},h_{\times}]^T_\mathrm{GR}
\label{eq:h}
\end{equation}
where 
\begin{equation}
    {\mathcal S}= \hat{\mathcal M} \ \text{diag}(1,\Delta)\hat{\mathcal M}^{-1}\,,
    \label{eq:S}
\end{equation} 
\begin{equation}
\hat{\mathcal M} = \left[
\begin{array}{cc}
-\sin(2\PL) & \cos(2\PL) \\
\cos(2\PL) & \sin(2\PL)
\end{array}\right]\,, 
\label{eq:mix}
\end{equation}
and $\Delta=e^{-i\omega \Delta t_{12}}$ 
with $\DT$ is the time delay between the polarisations and $\PL$ as the angle between the lens and the source, relative to the direction of GWs propagation that dictates the polarisation mixing. \\

It is easy to note that for $\PL = \pi/2 $,  $\mathcal{S} = \mathrm{diag}(1,\Delta)$, hence the signal observed by the detectors will just be superposition of $(+,\times)$ arriving at different times. 
\begin{equation}
h_I^{\MG}(t) = F^{+}_I h_{+}(t) + F^{\times}_I h_{\times}(t-\DT)
\label{eq:hmodGRsimp}
\end{equation}
whereas, if $\DT = 0$ the $\MG$  waveform morphology will be identical to the $\GR$ one, independent of $\PL$. 
Fig.~\ref{fig:mg_waveform} compares $\GR$ v/s $\MG$ waveform polarisations and the detector strains for various values $\DT$. 
Under $\MG$, the polarisations interfere leading to waveform distortions. 

Since lensing is an environmental effect which can occur through any local inhomogeneity in the path of GWs, the parameters $\Delta t_{12}$ and $\phi_{\rm lens}$ are expected to vary between GW events. The time delay distribution depends on the theory and the (usually unknown) lens properties and the configuration relative to the source. In general, one can only predict the probability of the birefringence parameters given a gravitational theory and matter distribution (unless further information or assumptions are employed about the source's location or the signal's trajectory), see Sec.~\ref{sec:cross_section_pheno}. This is in stark contrast to other tests of GW propagation (that are done with individual GW events) in which deviations represent a fundamental property of gravity (eg. massive graviton dispersion relations) and are thus the same across all events and only depend on their distance \cite{Will:2014kxa}. 

\subsection{Template Mismatch Studies}\label{sec:mismatch}

In order to quantify distortions due to GW birefringence, we calculate the mismatch between the $\GR$ and $\MG$ waveforms as seen the LIGO-Virgo detectors. At each detector ($I$), the mismatch between the injected waveform ($h_I^\mathrm{inj}$) and the recovery waveform ($h_I^\mathrm{rec}$)  is given by:
    \begin{equation}
    M_I = 1- \frac{( h_I^\mathrm{inj} | h_I^\mathrm{rec} )}{||h_I^\mathrm{inj}||.||h_I^\mathrm{rec}||}
    \end{equation}
where, $(\cdot\mid\cdot)$ symbolises the noise-weighted inner product:
\begin{equation}
(a~|~b)  \equiv 2\int_{f_{\mathrm{min}}}^{f_{\mathrm{max}}}\frac{\tilde{a}(f)\tilde{b}^{*}(f)}{S_n(f)}df\,.
\end{equation}
Here, $\tilde{a}, \tilde{b}$ represent the Fourier transform of the time series $a(t), b(t)$; $[f_{\mathrm{min}}, f_{\mathrm{max}}]$ is the frequency range over which the inner product is evaluated;  $^*$ represents complex conjugation and $S_n(f)$ is the colored Gaussian noise power spectral density (PSD) at the detector. The norm $||h|| = \sqrt{(h|h)}$ is the optimal SNR of a waveform.
We define the total mismatch ($ \mathcal{M} $) for network of detectors as the signal-to-noise ratio (SNR, $\rho_I$) squared weighted average of individual detector matches,
    
    \begin{equation}
    \mathcal{M} = \frac{\sum_{I} \rho_{I}^2 M_I}{\sum_I \rho_{I}^2}
    \end{equation}
    
Note that the mismatch is a normalized quantity and is maximised over time and phase shifts. Thus, the mismatch quantifies differences in morphology between the signals.
Whereas, during the parameter estimation (PE) from GW signals both the mismatch and SNR play a role as the log-likelihood  $\propto \mathcal{M}{\sum_I \rho_{I}^2}$ around the maximum likelihood parameters. We first wish to quantify the overall detectability of the birefringence. Later, we will estimate parameters using Bayesian inference, accounting for correlations between all parameters.
\begin{figure}[tbh]
    \centering
    \includegraphics[width = 0.98 \linewidth]{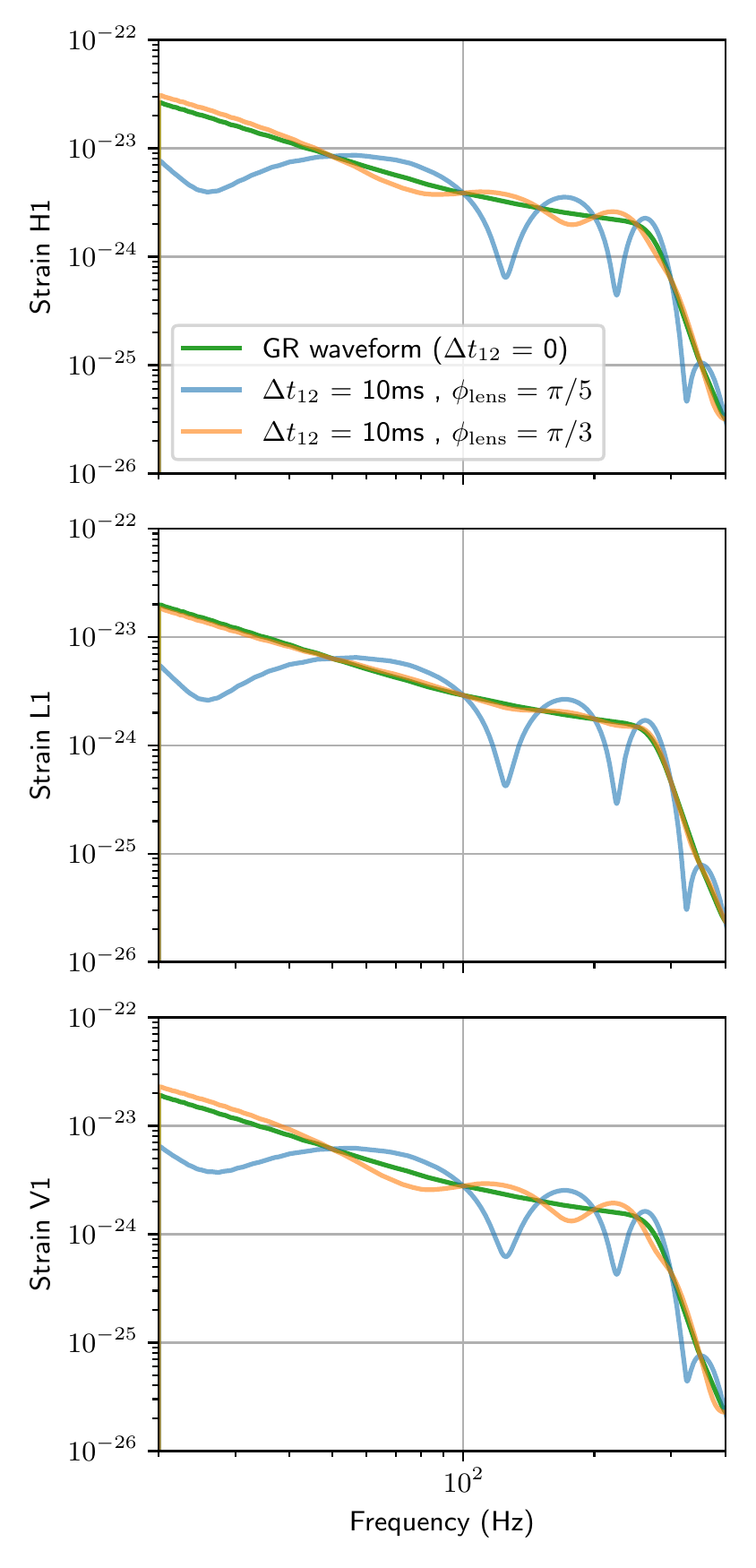}
    \caption{$\GR$ and $\MG$ detector frame waveform  amplitudes in frequency domain of GW150914-like CBC. The birefringence leads to additional frequency modulations and distorts the $\GR$ waveform. The magnitude of these distortions are however dependent on the two parameters: $\DT $ and $\PL$. }
    \label{fig:waveform}
\end{figure} 

\begin{figure*}[tbh]
    \includegraphics[width = 0.49 \linewidth]{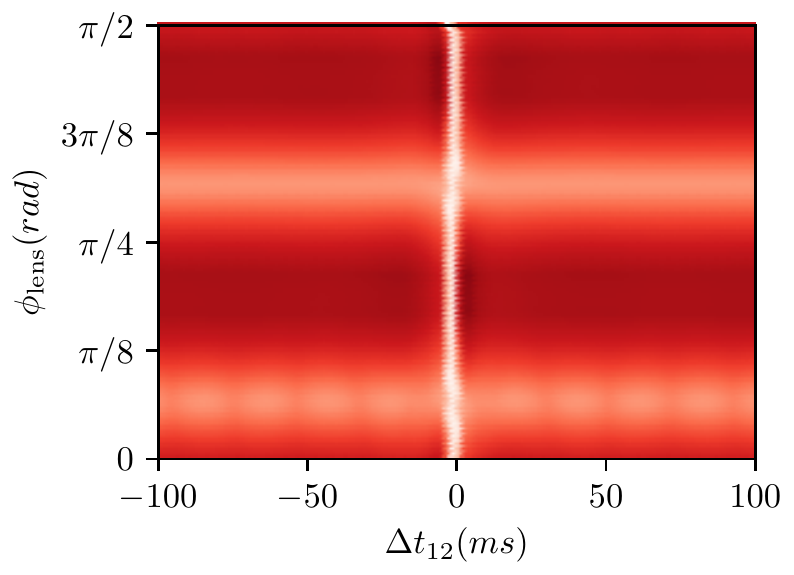}
    \includegraphics[width = 0.49 \linewidth]{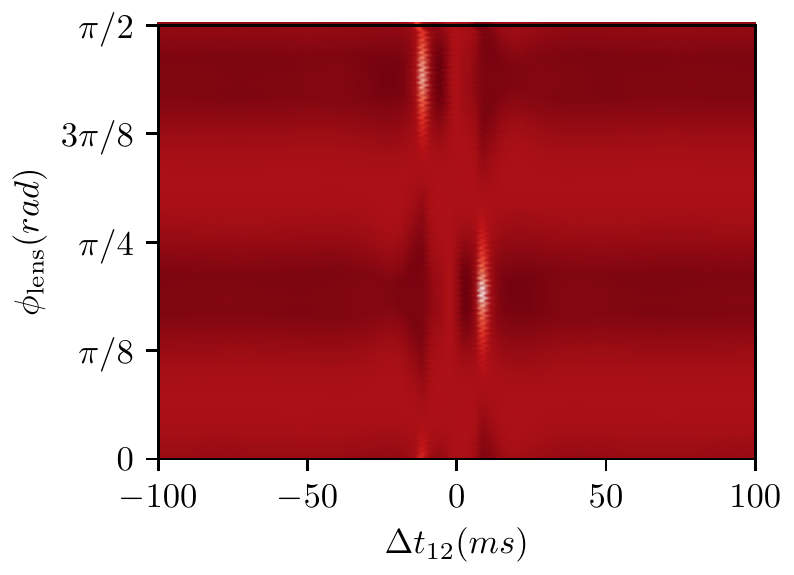}
    \includegraphics[width = 0.49 \linewidth]{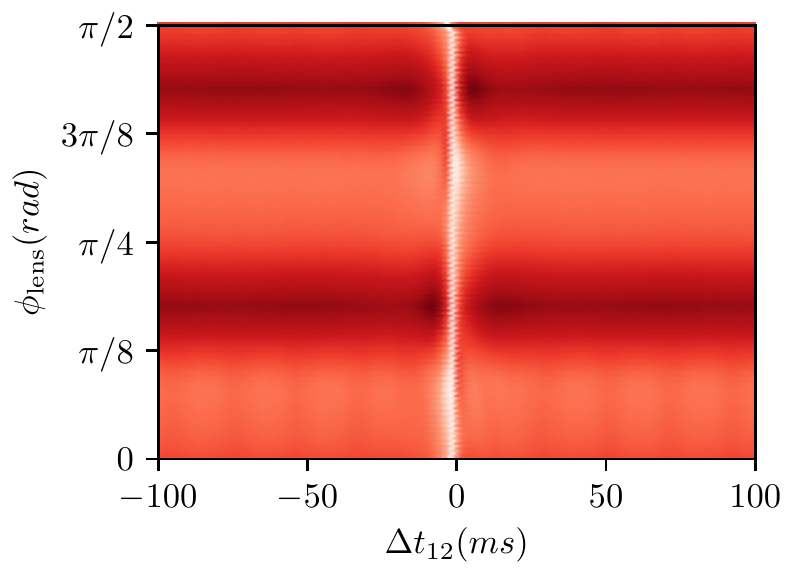}
    \includegraphics[width = 0.49 \linewidth]{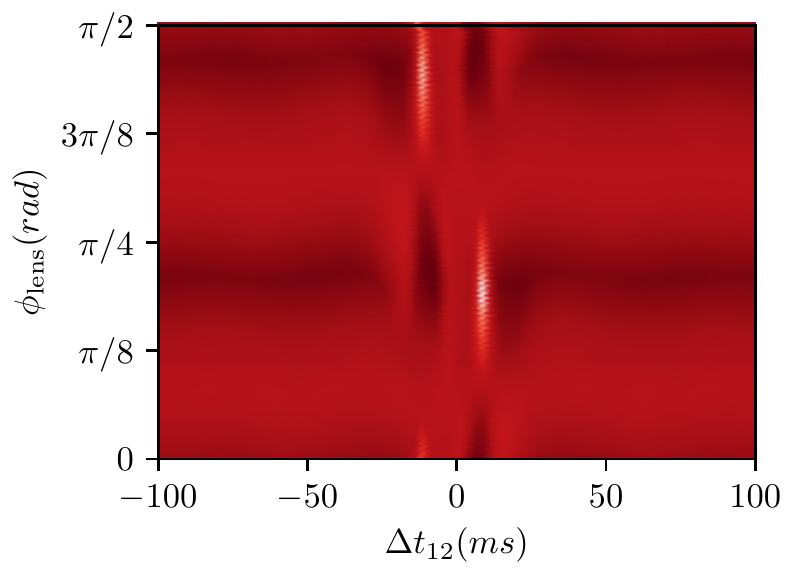}\\
    \includegraphics[width = 0.49 \linewidth]{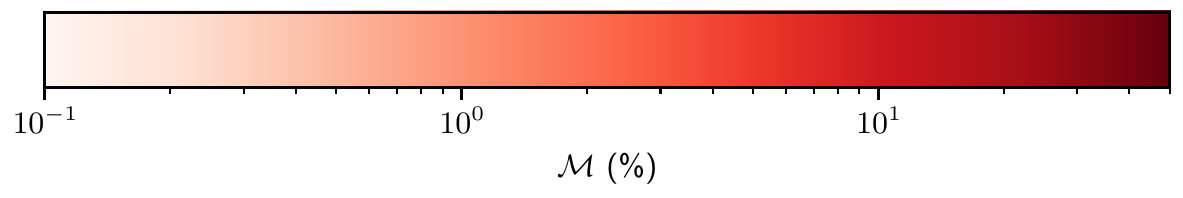}
    \caption{Mismatch between $\GR$ and $\MG$ waveforms for GW150914-like CBC (top) and GW190814-like CBC (bottom). left pannel: $\GR$ injection i.e. $\DT^\mathrm{inj}=0$ and $\phi_\mathrm{lens}^\mathrm{inj}=0$.  the mismatch is minimum for  $\DT^\mathrm{rec} \simeq 0 $. right pannel: a $\MG$ injection with $\DT^\mathrm{inj}=10 $ ms  and $\phi_\mathrm{lens}^\mathrm{inj}=\pi/5$.The mismatch is minimum at $\DT^\mathrm{rec} \simeq \pm 10 \mathrm{ms}$ and $\phi_\mathrm{lens}^\mathrm{rec} \simeq \pi/5, \pi/4 +\pi/5$.}
    \label{fig:mismatch}
    \end{figure*}

Fig.~\ref{fig:waveform} shows frequency domain $\MG$ and $\GR$ waveforms for a GW150914-like CBC. The waveforms are generated using the approximant $\textsc{IMRPhenomXPHM}$ \cite{IMRPhenomXPHM}, as implemented in the $\textsc{LALSimulation}$ module of the $\textsc{LALSuite}$ software package~\cite{lalsuite}. 
The waveforms are then projected onto the LIGO and Virgo detectors using their antenna pattern functions, as implemented in the $\textsc{bilby}$ \cite{bilby} software package. The $\MG$  waveforms have  additional frequency modulations which depend on the two parameters: $\DT $ and $\PL$ (see~\ref{sec:phenom_model}). Therefore we calculate the mismatch between the $\GR$ and $\MG$ waveforms, keeping all the other parameters identical and fixed for the two waveforms. 
In practice, we calculate ${M_I}$ using  $\textsc{pycbc.filter}$ \cite{pycbc} module. 
The detector noise is generated using the zero-detuned high-power PSDs of Advanced LIGO and Advanced Virgo at their design sensitivities \cite{AVirgoSensitivity,aLIGOSensitivity}. 

We consider two systems of binaries, one whose parameters resemble to that of the first CBC detection GW150914 and other of a higher mass-ratio CBC GW190814 where the presence of higher-order modes (HoMs) of GWs are significant. For both the systems, we inject a $\GR$ and a $\MG$ waveform and recover with $\MG$ waveform to calculate the mismatch. The parameters for both the CBCs are mentioned in Appendix~\ref{sec:app_inj}

Fig.~\ref{fig:mismatch} shows mismatches for a GW150914-like CBC (top) and GW190814-like CBC  (bottom). 
As expected for a $\GR$ injection i.e. $\DT^\mathrm{inj}=0$ and $\phi_\mathrm{lens}^\mathrm{inj}=0$, the mismatch is minimum for  $\DT^\mathrm{rec} \simeq 0 $, for all $\PL$ as expected from Eq. ~\ref{eq:S}. 
 Additionally, the local minimum of mismatch is at $\phi_\mathrm{lens}^\mathrm{rec} \simeq \pi/3 $, which could be because of vanishing polarisation ($+$ or $\times$) as seen at the detectors which further makes the mismatch independent of the time delay $\DT$. 
The waveform plots in Fig.~\ref{fig:waveform} confirms this as the $\phi_\mathrm{lens}^\mathrm{rec} \simeq \pi/3 $ waveform resembles the $\GR$ ones more as compared to the $\phi_\mathrm{lens}^\mathrm{rec} \simeq \pi/5 $, especially in the Livingston (L1) detector for  GW150914-like CBC. 

We also checked the mismatch for the $\MG$ injections (right panel Fig.~\ref{fig:mismatch}) with $\DT^\mathrm{inj}=10$ ms and $\phi_\mathrm{lens}^\mathrm{inj}=\pi/5$ and the mismatch is minimum at,  $\DT^\mathrm{rec}\simeq \pm 10$ ms and $\phi_\mathrm{lens}^\mathrm{rec} \simeq \pi/5, \pi/4 +\pi/5$. 
We can infer the degeneracy between $\DT$ and the coalescence time ($t_c$) as follows, 
from Eq.~\eqref{eq:ht}-\eqref{eq:mix} if $\PL \rightarrow \PL + \pi/4$ and $\Delta \rightarrow 1/\Delta$ then, one finds $S \rightarrow  S/\Delta$, which implies that the modification at $(\DT, \PL, t_c)$ is the same as at $(-\DT, \PL +\pi/4,t_c + \DT)$. Note that, this degeneracy stems from the fact that we do not know the composition of $h_{+,\times}$ before it encounters the lens. Additional information about the source could break this degeneracy, but we leave these investigations for the future.   

\subsection{Bayesian Inference}\label{sec:Bayes}

 Bayesian model selection allows us to assign evidences for various hypotheses pertaining to the observed data, and also derive posterior probability distributions of the model parameters conditioned on individual hypotheses. Given the set of data $\{d\}$ from a network of detectors, the marginalized likelihood (or, Bayesian evidence) of the hypothesis $\mathcal{H_A}$ can be computed by 
\begin{equation}
P(\{d\}|\mathcal{H_A}) = \int d\btheta P(\btheta|\mathcal{H_A}) \, P(\{d\}|\btheta, \mathcal{H_A}),
\label{eq:marg_likelihood}
\end{equation}
where $\btheta$ is a set of parameters that describe the signal under hypothesis $\mathcal{H_A}$ (including the masses and spins of the compact objects in the binary, location and orientation of the binary and the arrival time and phase of the signal), $P(\btheta|\mathcal{H_A})$ is the prior distribution of $\btheta$ under hypothesis $\mathcal{H_A}$, and $P(\{d\}|\btheta, \mathcal{H_A})$ is the likelihood of the data $\{d\}$, given the parameter vector $\btheta$ and hypothesis $\mathcal{H_A}$. Given the hypothesis $\mathcal{H_A}$ and data $\{d\}$, we can sample and marginalize the likelihood over the parameter space using an appropriate stochastic sampling technique such as nested sampling~\cite{skilling2006}. 

Bayesian model selection allows us to compare multiple hypotheses. For e.g., the odds ratio $\mathcal{O}^{\MG}_\mathrm{GR}$ is the ratio of the posterior probabilities of the two hypotheses $\MG$ and $\GR$. When $O^{\MG}_\mathrm{GR}$ is greater than one then hypothesis $\MG$ is preferred over $\GR$ and vice versa. Using Bayes theorem, the odds ratio can also be written as the product of the ratio of the prior odds $P^{\MG}_\mathrm{GR}$ of the hypotheses and the likelihood ratio, or Bayes factor $\BF$:

\begin{align}
\mathcal{O}^{\MG}_\mathrm{GR} : &= \frac{P(\mathcal{H}^{\MG}|\{d\})}{P(\mathcal{H}^{GR}|\{d\})} = \frac{P(\mathcal{H}^{\MG})}{P(\mathcal{H}^{GR})} \times \frac{P(\{d\}|\mathcal{H}^{\MG})}{P(\{d\}|\mathcal{H}^{GR})} \\ 
& =  \mathcal{P}^{\MG}_\mathrm{GR} \times \BF
\end{align}

Since $\GR$ has been tested well in a variety of settings, our prior odds are going to be highly biased towards it, i.e., $\mathcal{P}^{\MG}_\mathrm{GR} \ll 1$. Hence, in order to claim evidence of birefringence the corresponding Bayes factor supporting the $\MG$ hypothesis has to be very large. Since the Bayes factor is the only quantity that is derived from data, for the rest of the paper, we focus on the Bayes factor, i.e. the ratio of evidences under the two hypothesis.

The waveforms under the $\GR$ and $\MG$ hypotheses at each detector is the same way as described in Sec.~\ref{sec:mismatch}. 
We use the standard Gaussian likelihood model for estimating the posteriors of the parameters under different hypotheses~(see, e.g.,~\cite{Veitch:2014wba}). 
We use uniform priors in redshifted component masses of the binary, isotropic sky location (uniform in $\alpha, \sin \delta$) and orientation (uniform in $\cos \iota, \phi_0$), uniform in polarisation angle $\psi$, and a prior $\propto d_L^2$ on luminosity distance. Additionally for the $\MG$ hypothesis,  we choose the priors on $\DT$ as uniform $\in [-100  , 100 ]$ ms   and $\phi_\mathrm{lens}$ as uniform $\in [0,\pi/2] $. To estimate the posterior distribution and evidences for the $\GR$ and $\MG$ hypotheses, we use the open-source parameter estimation package \textsc{bilby} package \cite{bilby} coupled with the dynamical nested sampler \textsc{dynesty}~\cite{Dynesty:2020}. 

\subsection{Lensing Probabilities}\label{sec:cross_section_pheno}

The (non-)observation of birefringence can help us put constraints on theories beyond $\GR$ that predict $\MG$. According to \GR, the strong lensing of GWs caused by galaxies occurs when sources lie inside the Einstein radius of the lens,  
which depends on lens mass and profile. This is the relevant scale determining the probability of lensing. 
However, birefringence beyond $\GR$ is in principle independent of the ratio between the impact parameter and the Einstein radius, changing the probability of observing $\MG$ compared to strong lensing. 
It is thus possible to have $\MG$  time delays without multiple images, but birefringence could also occur for strongly lensed GWs, in which case it applies to each image separately, as typical time delays between images will be larger than $\Delta t_{12}$ \cite{Ezquiaga:2020dao}.

Assuming that the lenses are randomly distributed, birefringence detection is described by Poisson statistics.
A series of observations with $L$ lensed and $U$ unlensed GW events has an associated probability,
\begin{equation}
    P=\exp\Big({-\sum_i^U\lambda_{i}}\Big)\prod_j^L\left(1-e^{-\lambda_{j}}\right)\,.
\end{equation}
 The result depends on  $\MG$ rate for the $i^{th}$ event: $\lambda_i = \int dz_s d\vec p_L d\vec p_S \tau(z_s, \vec p_L) \mathcal{P}_i(z_s,\vec p_S)P(\vec p_S,\vec p_L)$. 
Here $S,L$ denote parameters corresponding to the source and lens/theory (i.e. beyond GR), $\mathcal{P}_i$ is the posterior distribution of the source parameters and $P$ is the prior, which includes relations between parameters (i.e. the measured $\DT$ as a function of lens mass and beyond-$\GR$ parameters).%
\footnote{A more complete treatment should account for the selection function \cite{Vitale:2020aaz}. In terms of gravitational lensing, we expect that $\DT$ correlates with magnification especially at sizeable impact parameters (e.g. single image regime, first magnified image if multiple images are formed), which dominate the lensing cross-section
Then events with larger $\DT$ are more likely to be observed and neglecting this correlation is conservative.}
The \textit{birefringence optical depth}, $\tau(z_s,\vec p_L)$ is the fraction of the sky for which $\MG$ is detectable for sources at a redshift $z_s$. 
Hereafter we will assume the posterior to be sharply peaked at the mean source redshift $z_s$ and include the integration on the lens model parameters ($\vec p_L$) in the definition of the optical depth, so $\lambda_i \approx \tau(z_{s,i})$. If birefringence is excluded in all events, the probability only depends on the total optical depth $\tau_{tot} \approx \sum_{i=1..N} \tau_i \approx \sum_{i=1..N} \lambda_i$.  
In Appendix~\ref{sec:beyond_poisson_stats} we will comment on opportunities to study GW birefringence beyond the Poisson statistics.

The lensing optical depth $\tau\left(z_{s}\right)$ depends on the angular cross section $\hat{\sigma} \left(z_{s}, \vec p_L \right)$ and the density of lenses $\hat n\left(\vec p_L \right)$  \cite{Xu_2022}.
In the following we will explicitly write the lens redshift $z_L$ and let $\vec p_{L^{\prime}}$ denote the remaining properties (i.e. lens mass \& theory parameters).
The total density of lenses at redshift $z_{L}$ is then $\int \hat n\left(z_{L}, \vec p_{L^{\prime}}\right) d \vec p_{L^{\prime}}$. The optical depth is computed directly by adding-up the cross-sections weighted by the density at different redshifts, i.e.
\begin{equation}
\tau\left(z_{s}\right)=\int_0^{z_{s}}d z_L \int d\vec p_{L^{\prime}}\frac{d V_c}{\delta \Omega d z_{L}} \hat n\left(z_{L},  \vec p_{L^{\prime}}\right) \hat{\sigma}\left(z_{s}, z_{L}, \vec p_{L^{\prime}}\right)
\label{eq:tau_z}
\end{equation}

where $d V_c =\delta \Omega D_{L}^2 \frac{dz} {(1+z) H(z)}$ is the physical volume given the solid angle $\delta \Omega$, angular diameter distance to the lens $D_{L}$  and  the Hubble parameter $H(z)$. For simplicity, we will assume point mass lenses of mass $M$ throughout. In GR, the lensing cross-section is  $\sigma = \pi \theta_E^2$,  where $\theta_E = R_E/D_L=(\frac{4 G M  D_{LS}}{ c^2 D_L D_S})^{1/2}$ is the Einstein angle, $D_S$ is the distance to the source from the earth and $D_{LS}$ is the distance between lens and source.

The relation between the $\MG$-time delays, the theory parameters and the configuration of the lensed system is complex (See Sec.~VI of Ref.~\cite{Ezquiaga:2020dao} for a worked-out example in a viable Horndeski theory). For this reason we will first consider two phenomenological models with generic
dependences of lensing cross section.
As a first example, we will assume that the relevant $\MG$ scale is proportional to the Einstein 
angle, $\theta^E_X=\alpha_X\theta_E$, so that the cross section becomes
\begin{equation}\label{eq:cross_section_rE}
\sigma^E_{X} = \pi \alpha_X^2 \theta_E^2\,.    
\end{equation}

Then the optical depth is given by Eq. (74) in Ref. \cite{Ezquiaga:2020dao} where the lenses have been assumed point-like. 

Under these assumptions, lensing probabilities are independent of the mass function.%
\footnote{The mass independence also appears for the strong-lensing cross section for a distribution of point lenses \cite{Pei93,Zumalacarregui:2017qqd}. This differs from strong-lensing cross-section for extended lenses, where the lens mass affects the formation of multiple images \cite{Xu_2022}.}

In our second example we assume that the relevant $\MG$ scale is given by a constant physical scale associated to each halo. Moreover, we assume that this scale depends on the halo mass as a power law. Accordingly, the cross-section reads,
\begin{equation}\label{eq:cross_section_pheno}
    \sigma_{\rm ph}=\pi \frac{R_{12}^2}{D_L^2} \left(\frac{M}{10^{12}M_\odot}\right)^{2n}\,.
\end{equation}
The scale $R_{12}$ fixes the probability of lensing for halos with $M=10^{12}M_\odot$, while $n$ allows us to extrapolate to different halo masses. Below we will discuss some cases of interest.

We now generalize the expression for the optical depth presented in Ref. \cite{Ezquiaga:2020dao} (Eq. 76) to include a realistic halo mass function. 
The optical depth from Eq.~\eqref{eq:cross_section_pheno} is given by
\begin{equation}\label{eq:tau_phys_halo}
    \tau^{\rm ph}(z_s,n) = \Omega_M h \left(\frac{R_{12}}{22\rm kpc}\right)^2 \hat \tau(z_s,n)\,,
\end{equation}
where

\begin{equation}\label{eq:cross_section_integrals}
\begin{multlined}
\hat \tau(z_s,n) = \int_0^{z_s}dz\frac{(1+z)^2}{H(z)/H_0} \int d\log(M) \\ \times \left(\frac{M}{10^{12}M_\odot}\right)^{2n-1}
{f(M,z)}.
\end{multlined}
\end{equation}

Here $f(M,z)=\frac{M^2}{\rho_0}\frac{d \hat n}{dM}$ is the scaled differential mass function (dimensionless) with $\rho_0$ as the matter density of universe at $z=0$. We will use the Tinker \textit{et al.} form~\cite{Tinker:2008ff} as implemented in the \textsc{Colossus} package~\cite{Diemer:2017bwl}. 
As our approach is phenomenological, we assume a Planck $\Lambda$CDM cosmology~\cite{Planck:2018vyg}. 
The true optical depth of a consistent $\MG$ model will typically depend more strongly on the theory parameters (e.g. entering Eq.~\eqref{eq:cross_section_pheno} via $R_{12}$) than on the precise values of $H(z), f(M,z)$ of the underlying $\MG$ cosmology, including the effects of deviations from $\GR$ in cosmological expansion and structure formation. This is the case for the example theory discussed in Sec.~\ref{sec:theory_connection}: GW lensing effects are orders of magnitude more sensitive than solar-system tests (cf. Fig.~\ref{fig:horndeski_constrained}), in turn more stringent than current cosmological observations \cite{Zumalacarregui:2020cjh,Alonso:2016suf} (for theories without a screening mechanism).

In addition to the Einstein radius scaling, Eq. \eqref{eq:cross_section_rE}, we will consider three cases of interest:
\begin{itemize}
    \item $n=1$: the physical scale is proportional to the total halo mass, much like the Schwarzschild radius. The rates are dominated by large masses and saturate at $z_s\gtrsim 1$, as the more massive halos are exponentially suppressed at early times. This case captures the dependence of the time delay in a Horndeski model (Sec.~\ref{sec:theory_connection}).
    \item $n=1/2$: the scale has the same mass scaling as the Einstein radius and leads to rates independent of $M$. However, the overall redshift dependence is different, as $R_E$ depends also on $D_S,D_{LS}$.
    \item $n=1/3$: this mass scaling favors lighter halos and thus grows very rapidly with redshift. It is motivated by the mass-dependence of the Vainshtein radius $R_V$, i.e. the classical strong-coupling scale \cite{Vainshtein:1972sx}. For $n=1/3$ the contribution from lighter halos diverges and a low mass cutoff needs to be included (we will take $M>10^7 M_\odot$). We will see this mass dependence when considering a binary merging near an active galactic nuclei in a Horndeski theory (Sec.~\ref{sec:theory_190521}).
\end{itemize}
The optical depths for each of the cases as a function of the source redshift are plotted in Fig.~\ref{fig:optical_depths_constrained}.
Note that these phenomenological models assume that the cross section is independent of $\DT$, and thus common for all the analyzed events. Dependence in the time delay can be included, e.g. by multiplying Eqs.~\eqref{eq:cross_section_rE},~\eqref{eq:cross_section_pheno} by a factor $(\DT/10{\rm \mathrm{ms}})^{-k}$. For the sake of simplicity, we will not include this dependence and instead interpret the obtained values of $\alpha_X$, $R_{12}(n)$ at the median 95\% c.l. from all analyzed events.

\section{Results}\label{sec:results}

In order to test our method and understand the observing capabilities, we first apply our pipeline to injections. We then proceed to analyse the latest GW catalog (GWTC-3).

\begin{figure*}[t!]
\centering

\subfigure[$SNR$ $10$]{\label{fig:a}
\includegraphics[width=0.32\linewidth]{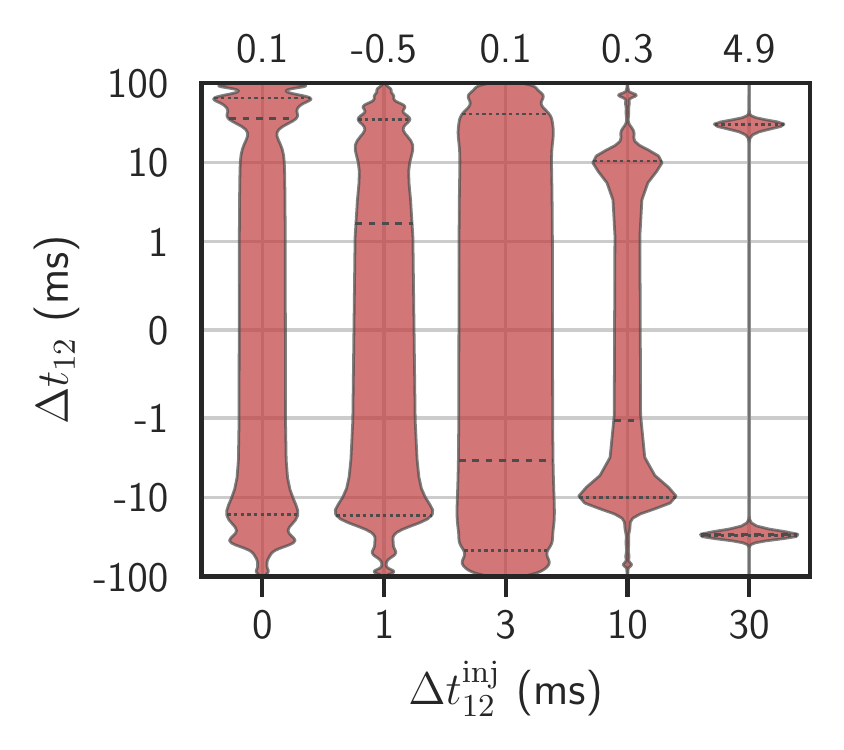}}
\subfigure[$SNR$ $15$]{\label{fig:b}
\includegraphics[width=0.32\linewidth]{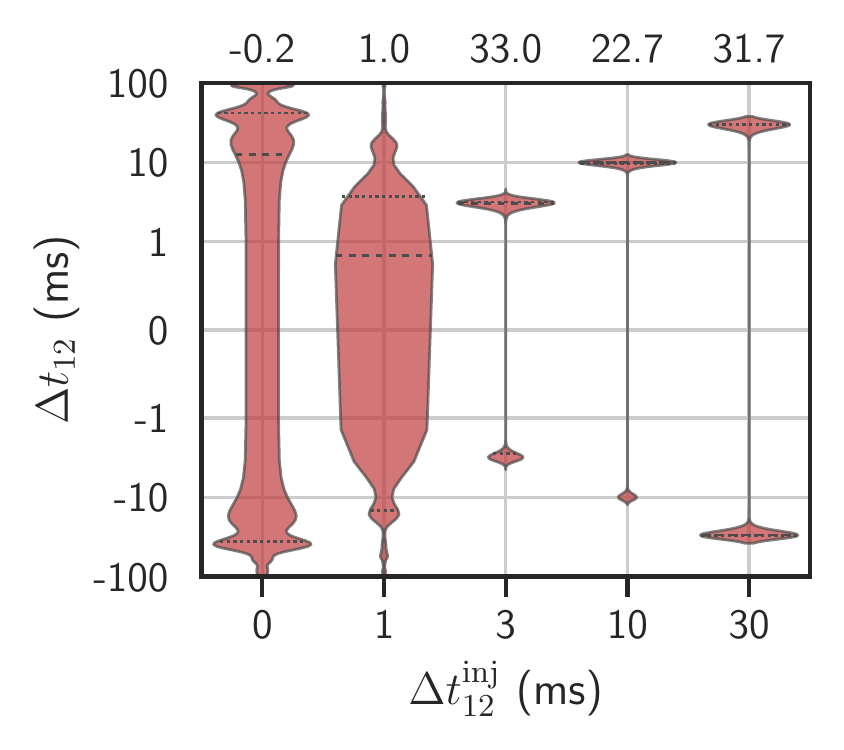}}
\subfigure[$SNR$ $20$]{\label{fig:c}
\includegraphics[width=0.32\linewidth]{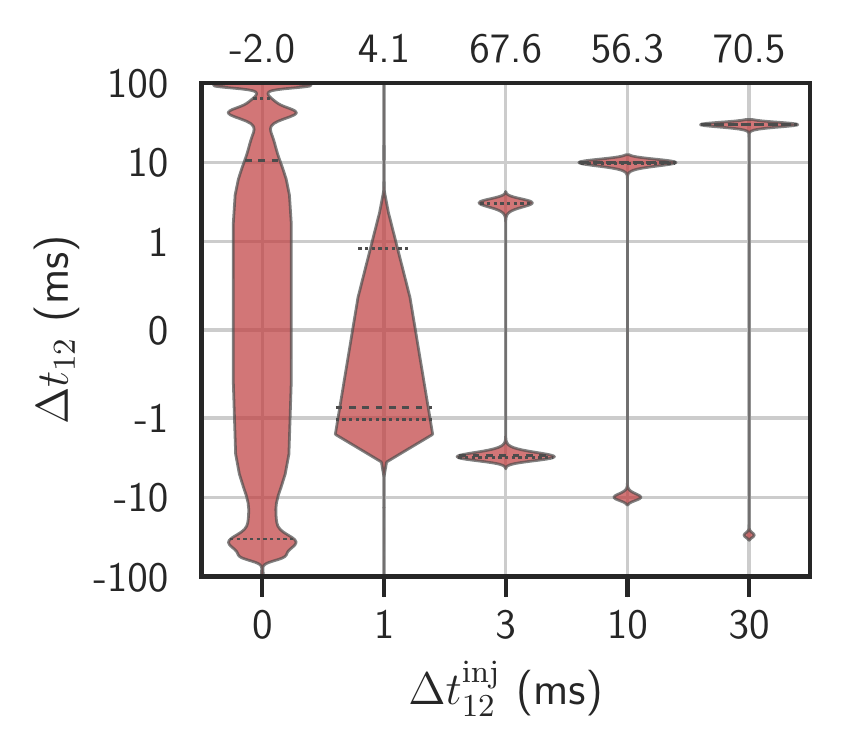}}
\subfigure[$SNR$ $30$]{\label{fig:d}
\includegraphics[width=0.33\linewidth]{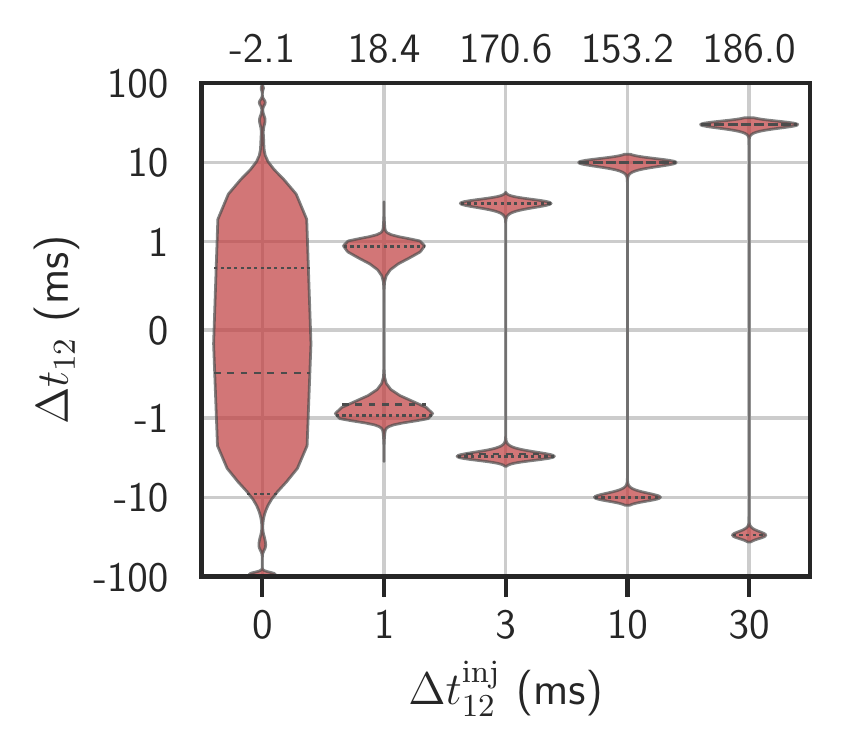}}
\subfigure[$SNR$ $40$]{\label{fig:e}
\includegraphics[width=0.33\linewidth]{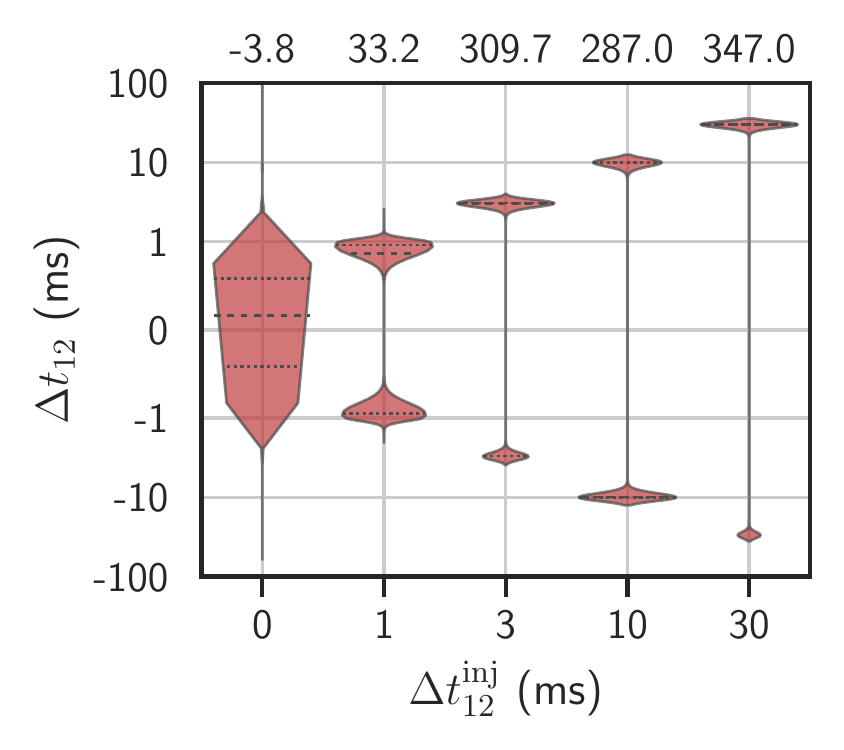}}
\caption{Signal-to-noise ratio (SNR) dependence of $\DT(\mathrm{ms})$ posteriors and the  $\log \BF$ (upper-x axis) for the GW150914-like injections with different values of $\DT^\mathrm{inj}$ (lower-x axis) and $\phi_\mathrm{lens}^\mathrm{inj} = \pi/5$.  Time delays ($\DT$) as small as $1ms$ are recovered well with SNR 30 \& 40 signals, and for SNR 10 signals time delays $<30ms$ are not measurable. Both model selection and time delay measurements (without the symmetry around $\DT=0$) improve with increase in SNR.}
\label{fig:sim_events}
\end{figure*}

\subsection{Injections}
We inject GW150914-like signals in simulated Gaussian noise with $\DT^\mathrm{inj} \in \{0, 1, 3, 10, 30\}$ ms and  $\phi_\mathrm{lens}^\mathrm{inj}= \pi/5$ rad and recover them by running the parameter estimation routines under the $\GR$ and $\MG$ hypothesis, as mentioned in Sec.~\ref{sec:Bayes}.  This allows us to compute the Bayes factors $\BF$ to compare the two hypothesis for each injection.

The injections are set to have $\mathrm{SNR} \in \{10, 15, 20, 30, 40\}$. These SNRs are achieved by inversely scaling the luminosity distance ($d_L$) of the injections. Fig.~\ref{fig:sim_events} shows the violin plots and the $\log$ Bayes Factors for these injections. 
The posteriors on  $\phi_\mathrm{lens}$ are uninformative in all the cases and hence not shown in the figure. 
$\MG$ time delays ($\DT$) as small as 1 ms are recovered well with SNR 30 and 40 signals, whereas for SNR 10 signals time delays $<30$ ms are not measurable. As one would expect, only with $\DT^\mathrm{inj} = 0$, i.e. $\GR$ injection the $\log \BF < 0 $, i.e. consistent with the $\GR$ hypothesis, except for the SNR 10 case where $\log \BF = 0.1 $ is  within the intrinsic sampling error on the calculation of evidence. For $\DT^\mathrm{inj} \in \{1, 3, 10, 30\}$ ms we find that $\log \BF > 0 $, i.e. consistent with the $\MG$ hypothesis for all the SNRs except 10. Hence, both model selection and sensitivity to measure the time delays improve with an increase in SNR. 
 
We also note that the time delays are measurable up to a  symmetry around $\DT=0$. This is because $\MG$ waveform, Eq. \eqref{eq:S}, is identical at $(\DT, \PL, t_c)$  and $(-\DT, \PL +\pi/4,t_c + \DT)$, which we also saw during mismatch studies with $\MG$ injections, see right panel of Fig.~~\ref{fig:mismatch}. 
It is possible that for asymmetric and inclined binaries with significant HoMs a better measurement of $\PL$  could break the $\DT$ parity as well, however, this needs to be investigated further and is left for future studies.

Overall, as the sensitivity of the detectors improve we shall be able to measure the birefringence time delays as small as $1$ ms. On the other hand, in the absence of birefringence we expect to see $\DT$(s) posteriors which are consistent with the $\GR$ value, i.e. zero and bayes factors that favour the $\GR$ hypothesis. Most events in GWTC-3 have SNR $<30$. The time delay posteriors are hence expected to be broad, however the Bayes factors should already indicate whether $\MG$ is present or not. 

\subsection{GWTC-3 Events}
We now analyse 43 CBC events from the GWTC-3, that have low detection false alarm rate, $\mathrm{FAR}\lesssim 10^{-3} \ \mathrm{yr}^{-1}$. These are also the events that are considered for other tests of $\GR$ performed previously~\cite{LSC_2016grtests,GWTC1TestofGR,LIGOScientific:2020tif, LIGOScientific:2021sio}.

\begin{figure*}[tbh]
    \centering
    \includegraphics[ width = \linewidth,clip ]{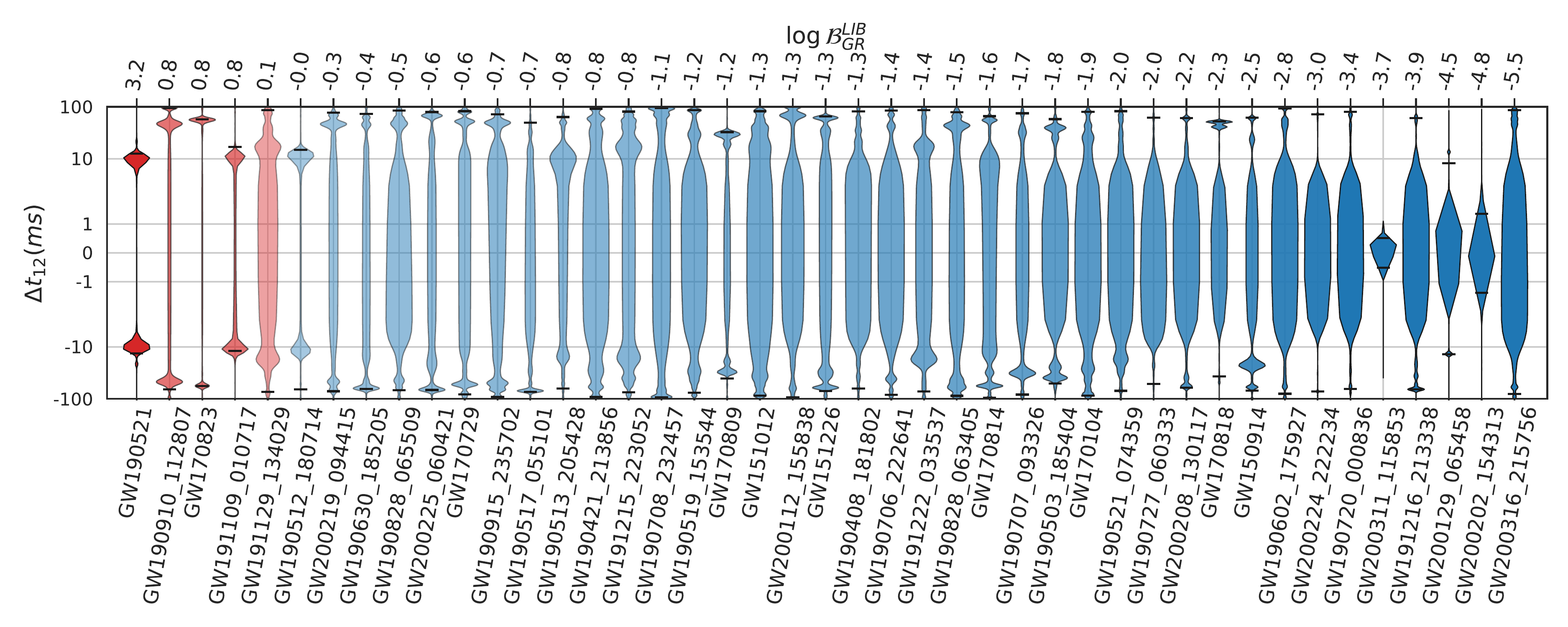}
\caption{lens-induced birefringence ($\MG$) test of GWTC-3 events \cite{LIGOScientific:2021djp}. We show the posteriors on $\DT(\mathrm{ms})$ and Bayes Factors $\log \BF$ (upper $x$-axis). Events with positive Bayes factors are highlighted in red.}
    \label{fig:real_events}
\end{figure*}

Fig.~\ref{fig:real_events} shows the $\DT$ posteriors and the $\log$ Bayes Factors for the real events. We find that for almost all the events the  $\DT$ posteriors are broad containing zero, i.e. consistent with $\GR$. This is mostly due to the low SNRs of the events, as seen in our injection study. We also find the tightest 90\% credible bounds on $|\DT| \lesssim 0.51$ ms coming from the event 
GW200311\_115853 which has reasonably high SNR ($\simeq 17.8$) and moderate redshift ($z\sim 0.23$) as compared to other events. As expected $\phi_\mathrm{lens}$ posteriors are uninformative for almost all the events.

38 out of 43 events resulted in $\log \BF < 0 $, and hence consistent with the $\GR$ hypothesis. Only a few events showed preference to $\MG$ hypothesis ($\log \BF >0$), with highest one for GW190521 (3.21)  and then  GW190910\_112807 (0.8), GW170823 (0.8), GW191109\_010717 (0.7) \& GW191129\_134029 (0.1).

\begin{figure}[h]
    \centering
    \includegraphics[width = \linewidth]{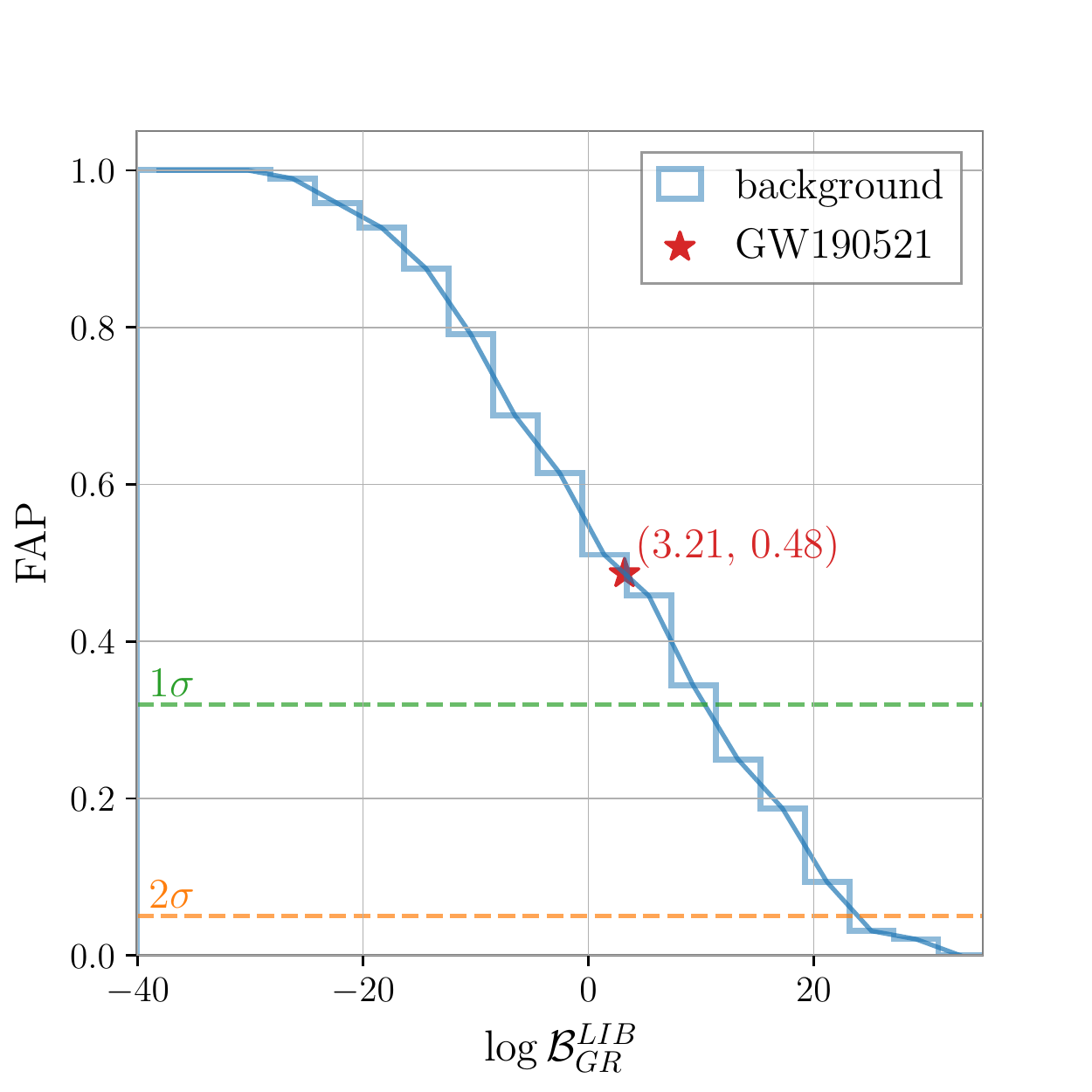}
    \caption{Bayes factors distributions for GW190521-like CBC, calculated by doing PE with both the hypothesis, for $\sim 100$ $\GR$ injections from the GW190521 posteriors in different realisations of gaussian noise. The false alarm probability for the observed $\log \BF = 3.2$ is found to be $0.48$. }
    \label{fig:190521_bg}
\end{figure}

The Bayes factors are known to be prior dependent and its value does not signify the confidence in preferring one hypothesis over the other, but rather the preference of one hypothesis over the other given a set of prior assumptions. 
The model with extra parameters ($\MG$) could be either fitting the noise or the signal, therefore we take a frequentist approach to determine the significance by considering different realisations of noise. We focus on the event with the highest Bayes factors (GW190521) and estimate its significance.
We generate the background distribution of Bayes factors by injecting $\GR$ signals in Gaussian noise using the power spectral density around the trigger time. To calculate the false alarm probability corresponding to the observed Bayes Factor for the event GW190521, we simulate a hundred $\GR$ injections, whose parameters are taken from the posteriors of GW190521 event for the $\GR$ hypothesis. Fig.~\ref{fig:190521_bg} shows the background distribution of the  Bayes factors
and the corresponding false alarm probability (FAP). The FAP corresponding to each $\BF = \kappa $ is calculated as the fraction of the background events having  $\BF > \kappa$. We find that for the observed $\log \BF = 3.2$ for GW190521  is $0.48$, i.e. its significance is less than $1 \sigma$.

It is to be noted that GW190521 is a remarkably loud but short ($ < 100 $ ms) signal,  being easily fit by widely different hypotheses such as head on-collision of a boson star \cite{CalderonBustillo:2020fyi} or left-right (L-R), frequency-dependent  birefringence \cite{Wang:2021gqm}. For the interested reader, in Appendix~\ref{sec:pe_190521} (Fig.~\ref{fig:GW190521}), we also plot posteriors and the waveforms corresponding to the maximum a posteriori  parameters for both the hypothesis, over the whitened signals observed at each detector. 
The plots show that the $\MG$ hypothesis is also fitting the noise, and might therefore give a high Bayes factor, $\BF$. The other events with $\BF > 0$ are also show similar behavior and as their preference for LIB is marginal, we conclude that none of the events have any significant Bayes factor and find \textit{no strong evidence} for birefringence. 

\section{Implications}
\label{sec: implications}
In our analysis of the latest GW catalog, we have found that the majority of the events disfavor birefringence. For a subset of them (most notably GW190521) while the Bayesian inference prefers the $\MG$ hypothesis, a follow-up background study indicates that most simulated $\GR$ signals give comparable Bayes factors.
In the following, we present the implications of these results. 
First, we consider the implications for generic $\MG$. Then, we study the constraints on a specific scalar-tensor theory that predicts $\MG$. 
Finally, we  entertain the possibility that GW190521 was emitted in an active galactic nucleus (AGN) and is displaying evidence of birefringence.
\subsection{Constraints on generic $\MG$}

From the non-observation of birefringence in the 43 events from GWTC-3 and using their median redshift values \cite{LIGOScientific:2021djp}, we estimate the total optical depth for the $\MG$ models discussed in Sec.~\ref{sec:cross_section_pheno}. The non-observation of birefringence translates to constraints on the phenomenological model parameters, as summarized in Table~\ref{tab:constraints}. For reference, we also show the constraints obtained from the full GWTC-3 (90 events).

\begin{table}[h]
    \centering
    \bgroup
\def\arraystretch{1.5}
    \begin{tabular}{l l   c}
$\sigma_{\rm \MG}^{1/2}\hspace{.2cm} $ \hspace{0.5cm}& 95\% c.l. & comment\\ \hline
$ \propto M$ & $R_{12} < 4.4\,(2.9)$ kpc &  Sec.~\ref{sec:theory_connection} \\ \hline
$\propto R_E$ &  $\alpha_E < 3.0\,(1.6)$ &  \\
$\propto M^{1/2}$ & $R_{12} < 20\,(12)$ kpc & \\\hline
$\propto M^{1/3}$   & $R_{12} < 12\,(6.9)$ kpc &\quad  \\\hline
    \end{tabular}
    \egroup
    \caption{Constraints on the phenomenological models Eq.~\eqref{eq:cross_section_rE} and Eq.~\eqref{eq:cross_section_pheno}, assuming no birefringence detected for analysed (all) GWTC-3 events. 
    \label{tab:constraints}}
\end{table}

\begin{figure}[tbh]
    \centering
    \includegraphics[width = \linewidth]{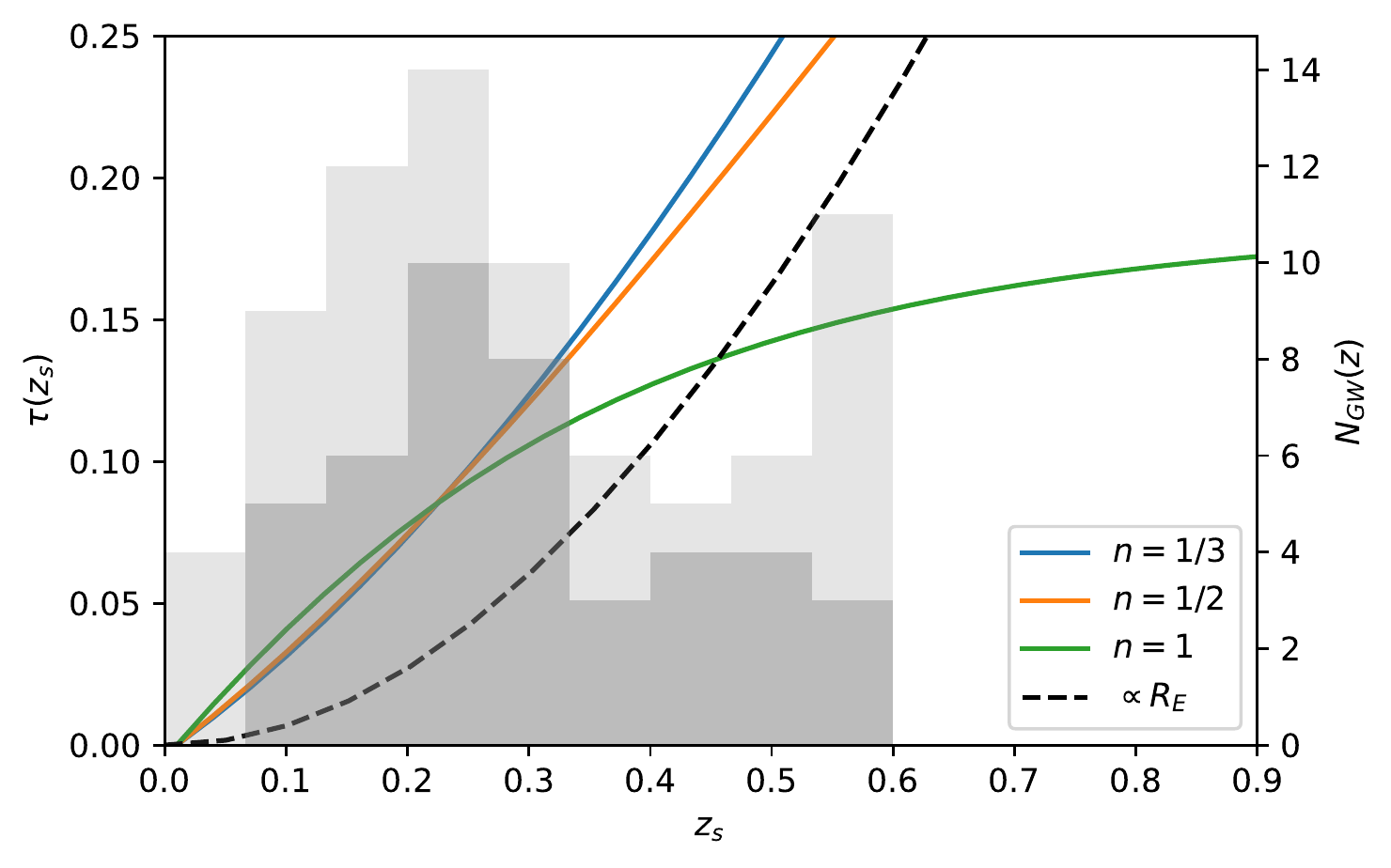}
    \caption{Birefringence optical depth for the phenomenological models considered here, using the parameters correspond to the 95\% c.l. limit compatible with the non-observation of $\MG$. The dark (light) gray shaded histograms show the binned redshift distribution of analysed (all) GWTC-3 events. See Sec.~\ref{sec:cross_section_pheno} for details.}
    \label{fig:optical_depths_constrained}
\end{figure}

The higher redshift events have higher optical depth. Non-observation of birefringence in distant sources leads to more stringent constraints, although the SNR scales with the inverse luminosity distance: hence some of the highest redshift events will not be considered because of our FAR threshold. The final results depend strongly on the model via source redshift and halo mass function. Figure~\ref{fig:optical_depths_constrained} shows the redshift dependence of the optical depth for the parameterizations discussed, adopting the 95\% c.l. values found by our analysis along with the observed GWTC-3 redshift distribution.

Future observations will increase in number of events and their SNRs, allowing better constrain the birefringence probabilities and ruling out more of the parameter space in the alternative theories of gravity. Higher-redshift observations above our FAR threshold will be especially valuable to constrain $\alpha_E$ and $R_{12}$ for $n=1/3,1/2$ (see Fig.~\ref{fig:optical_depths_constrained}).

\subsection{GW birefringence in Horndeski theories}\label{sec:theory_connection}

Let us now use our results to a specific theory that predicts $\MG$. We will present the theory and translate the constraints of the phenomenological model (Table~\ref{tab:constraints}) into fundamental theory parameters. In the next subsection we will interpret a tentative detection of $\MG$ in GW190521 as an AGN binary within the same theory.
We will focus on a particular scalar-tensor theory within the Horndeski class~\cite{Horndeski:1974wa}, whose $\MG$ predictions have been analyzed in detail, cf. Sec. 6 in Ref.~\cite{Ezquiaga:2020dao}.
The model is described by two parameters describing couplings between the Ricci scalar ($R$) and the new field $\phi$: a linear coupling $p_{4\phi}$ and a derivative coupling suppressed by an energy scale $\Lambda_4$. 
The Lagrangian of this theory can be written as \cite{Ezquiaga:2016nqo,Ezquiaga:2017ner}
\begin{equation}\label{eq:lagrangian_example_horndeski}
    \mathcal{L}\sim -\frac{1}{2}(\nabla\phi)^2+ \frac{M_P^2}{2}\left(1+\frac{p_{4\phi}\phi}{M_P}\right)R + \frac{\phi}{\Lambda_4^2}\nabla_\mu\nabla_{\nu}\phi G^{\mu\nu}\,,
\end{equation}
where $R$ is the Ricci scalar, $G_{\mu\nu}$ is the Einstein tensor, $M_P$ is the Planck mass in units of $c = h = 1 $,  and $\nabla$ the covariant derivative.
The $\GR$ limit corresponds to $p_{4\phi}\to 0, \Lambda_4\to \infty$ . 
The parameters of this model are stringently constrained by the speed of GWs on the homogeneous FRW metric \cite{Ezquiaga:2017ekz,Creminelli:2017sry,Baker:2017hug,Sakstein:2017xjx} (see also \cite{Brax:2015dma,Lombriser:2015sxa,Bettoni:2016mij}), as observed by the near-coincident arrival of GW170817 and its associated counterpart~\cite{Monitor:2017mdv}: $|c_g/c-1|\lesssim 10^{-15}$. Compliance with this limit requires \cite{Ezquiaga:2020dao}
\begin{equation}\label{eq:horndeski_170817}
p_{4\phi} \lesssim 10^{-8}\Lambda_4/H_0 \quad \text{(GW170817)}\,.
\end{equation}
While this constraint is extremely stringent, $\MG$ allows comparable limits.

Specifying a model allows one to derive concrete predictions. The dependence of the time delay contributions (Shapiro, geometric) with the lens and theory parameters is complex. Nonetheless, we observed that the time delay decreases monotonically with the impact parameter. Moreover, its slope changes and becomes very sharp beyond the Vainshtein radius
\begin{equation}\label{eq:r_vainshtein}
    r_V 
 = 1.2{\rm Mpc}\, p_{4\phi}^{1/3}\left(\frac{M}{10^{12}M_\odot}\right)^{1/3}\left(\frac{H_0}{\Lambda_4}\right)^{2/3}
    \,.
\end{equation}
$r_V$ represents the scale at which the scalar field has a strong self-coupling near a massive object \cite{Vainshtein:1972sx}\footnote{For extended lenses one needs to consider the effective Vainshtein radius, such that $r_V(M(r_V^{\rm eff})=r_V^{\rm eff}$, (see Eq. (186) and Fig. 14 in Ref. \cite{Ezquiaga:2020dao}).}.
In many scalar-tensor theories this leads to \emph{screening}: a suppression of scalar field fluctuations for $r<r_V$, allowing the theory to approximately recover $\GR$ around massive bodies. However, screening is not necessary in this model given the stringent constraint from GW170817 (Eq.~\eqref{eq:horndeski_170817}). In this case, the strong-interaction within $r_V$ represents a a large coupling between the scalar field and the Riemann tensor, the kind of interaction producing $\MG$.

For simplicity, we will focus on the Shapiro time delay. The geometric time delay is usually dominant for massive halos at intermediate distances. (Fig 12 in Ref.~\cite{Ezquiaga:2020dao}). It is proportional to the Einstein radius, and it could thus be captured generalizing Eq.~\ref{eq:cross_section_rE} to extended lenses. Neglecting the geometric time delay is conservative but reasonable, since our constraints involve events at relatively low redshift ($z\lesssim 0.6$).

The $\MG$ predictions have a simple dependence on the lens mass and theory parameters. We verified that $\Delta t_{12}\propto M \Lambda_4^{-4/3}$. The proportionality to the mass stems from the scaling with the Vainstein radius, as well as $\Delta t_{12}$, the impact parameter and the time spent by the GW on the region of sizeable birefringence are all $\propto r_V$. It allows us to directly connect the theory parameters to $R_{12}$ with $n=1$, as constrained in the phenomenological model \eqref{eq:cross_section_pheno}. The scaling with $\Lambda_4$ allows us then to find $R_{12}$ by equating $\Delta t_{12}(R_{12})$ to the constrained value for different $p_{4\phi}$, but keeping $M=10^{12}M_\odot$, $\Lambda_4$ fixed. 
For simplicity, we will take a sensitivity of $\Delta t_{12}\sim 10 $ ms to define $R_{12}$.
Using the actual posteriors on $\Delta t_{12}$ for each of the GWTC-3 events analyzed will not qualitatively affect these constraints in any significant manner.

\begin{figure}[tbh]
    \centering
    \includegraphics[width = \linewidth]{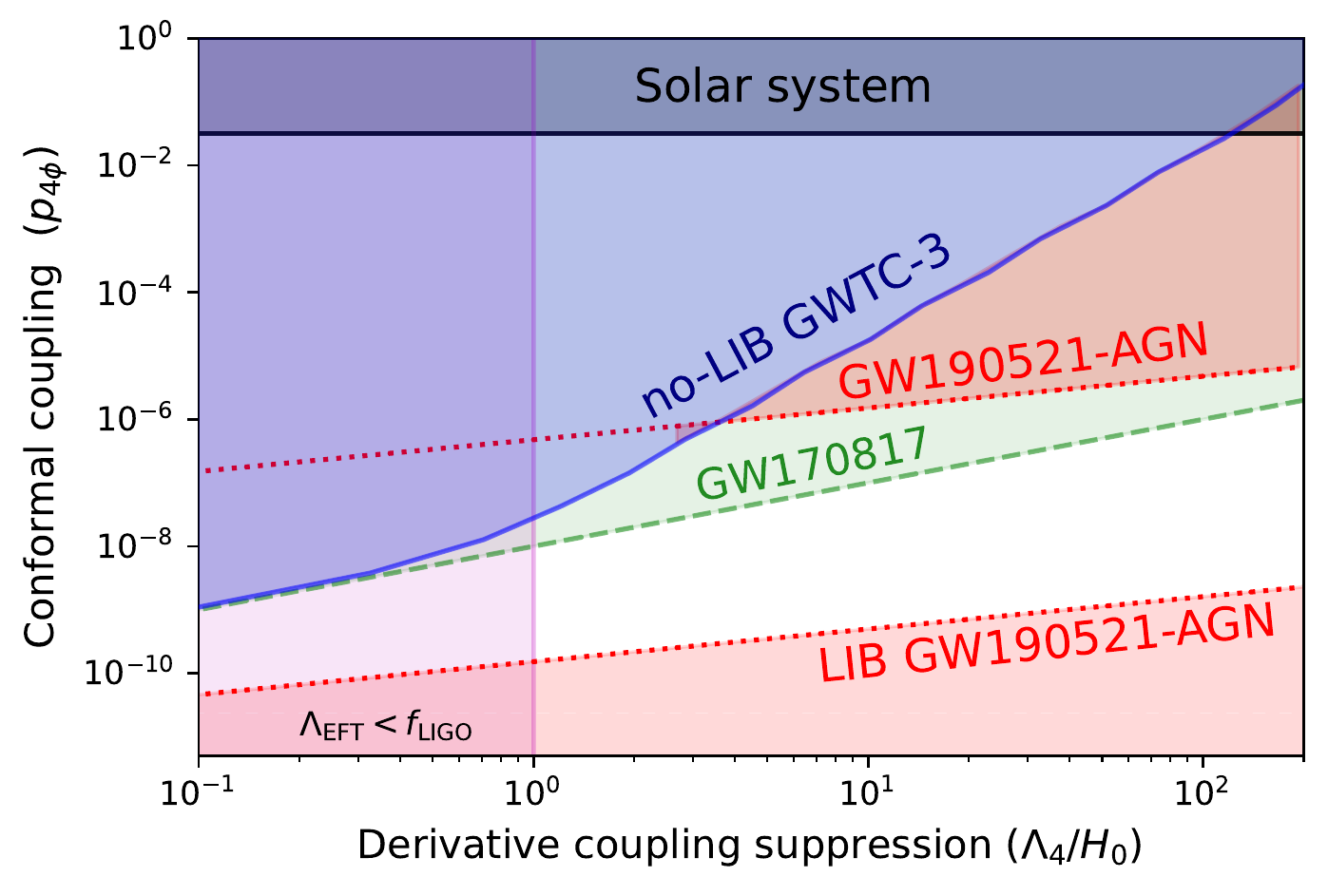}
    \caption{95\% c.l. constraints on the parameters of a quartic Horndeski theory \cite{Ezquiaga:2020dao} using the lens-induced birrefringence ($\MG$) test. Shaded regions are excluded according to GWTC-3 (this work, blue solid), GW170817 \cite{Ezquiaga:2017ekz,Creminelli:2017sry,Baker:2017hug,Sakstein:2017xjx} (green dashed) and GW190521 assuming an AGN binary
    ~\cite{Graham:2020gwr} (red dotted, see Fig.~\ref{fig:dt_agn_horndeski}). 
    The $\GR$ limit corresponds to $p_{4\phi}\to0, \Lambda_4\to \infty$, when the scalar field is decoupled from gravity and its derivative interactions suppressed. 
    See sections~\ref{sec:theory_connection},~\ref{sec:theory_190521} for details. 
    If GW190521 is associated to an AGN, the upper shaded region improves the overall GWTC-3 constraints for $\Lambda_4\gtrsim 3 H_0$. 
    If we further assume a detection of $\MG$, then the bottom red shaded region excludes GR. 
    For reference, we also indicate Solar System constraints (gray horizontal) and the region there the GW frequencies at LIGO-Virgo detectors are larger than the (non-linear) energy scale of the effective field theory (magenta vertical).}
    \label{fig:horndeski_constrained}
\end{figure}

The excluded region is shown in Fig.~\ref{fig:horndeski_constrained}, along with constraints from the GW speed on FRW and lunar laser ranging (no screening, $p_{4\phi}\ll1$, see Sec. VBc in Ref. \cite{Zumalacarregui:2020cjh}). The change in slope at low $\Lambda$ corresponds to a transition in which $R_{12}$ surpasses the Vainsthein radius \eqref{eq:r_vainshtein}. For $\Lambda_4 \ll H_0$ the birefringence constraints approach those of the GW speed: this happens when $r_V$ is so large that most GWs are effectively behind a lens. Then the constraints are satisfied in the limit $c_{\rm GW}\to c$, equivalent to Eq. \eqref{eq:horndeski_170817}. For the sensitivity of GWTC-3 this happens for  $\Lambda_4\lesssim H_0$, where LVK frequencies lie beyond the validity of our framework as a classical effective field theory \cite{deRham:2018red}. At increasing $\Lambda_4$ the constraints degrade, since probability becomes very suppressed, Eq. \eqref{eq:r_vainshtein}.
For $\Lambda^4\gtrsim 10^2H_0$ solar system constraints become more efficient than birefringence.

\subsection{GW190521 as an AGN binary} \label{sec:theory_190521}

Let us now discuss the implications of a possible birefringence detection associated to GW190521. Given the constraints from the speed of GWs \eqref{eq:horndeski_170817} on our example Horndeski theory, the chances of birefringence being caused by a lens in the line of sight are very small. We will instead interpret our result, $\Delta t_{12}\gtrsim 9.5$ ms as due to an environmental effect near the source. We will follow the scenario outlined in Ref.~\cite{Graham:2020gwr}, where a candidate electromagnetic counterpart from an  AGN J124942.3 + 344929, observed 34 days after the GW signal, suggests that the binary merged in the environment of a supermassive black hole (SMBH).
Note that there are important uncertainties, both regarding the counterpart association (given large GW localization uncertainties \cite{Palmese:2021wcv}), and the significance of $\MG$ detection (given our analysis of random noise realizations, Fig.~\ref{fig:190521_bg}). This discussion is therefore not a statement on the status of $\GR$. Instead, it proves the potential of identifying environments of GW sources to test gravity theories.

\begin{figure}
    \centering
    \includegraphics[width=\columnwidth]{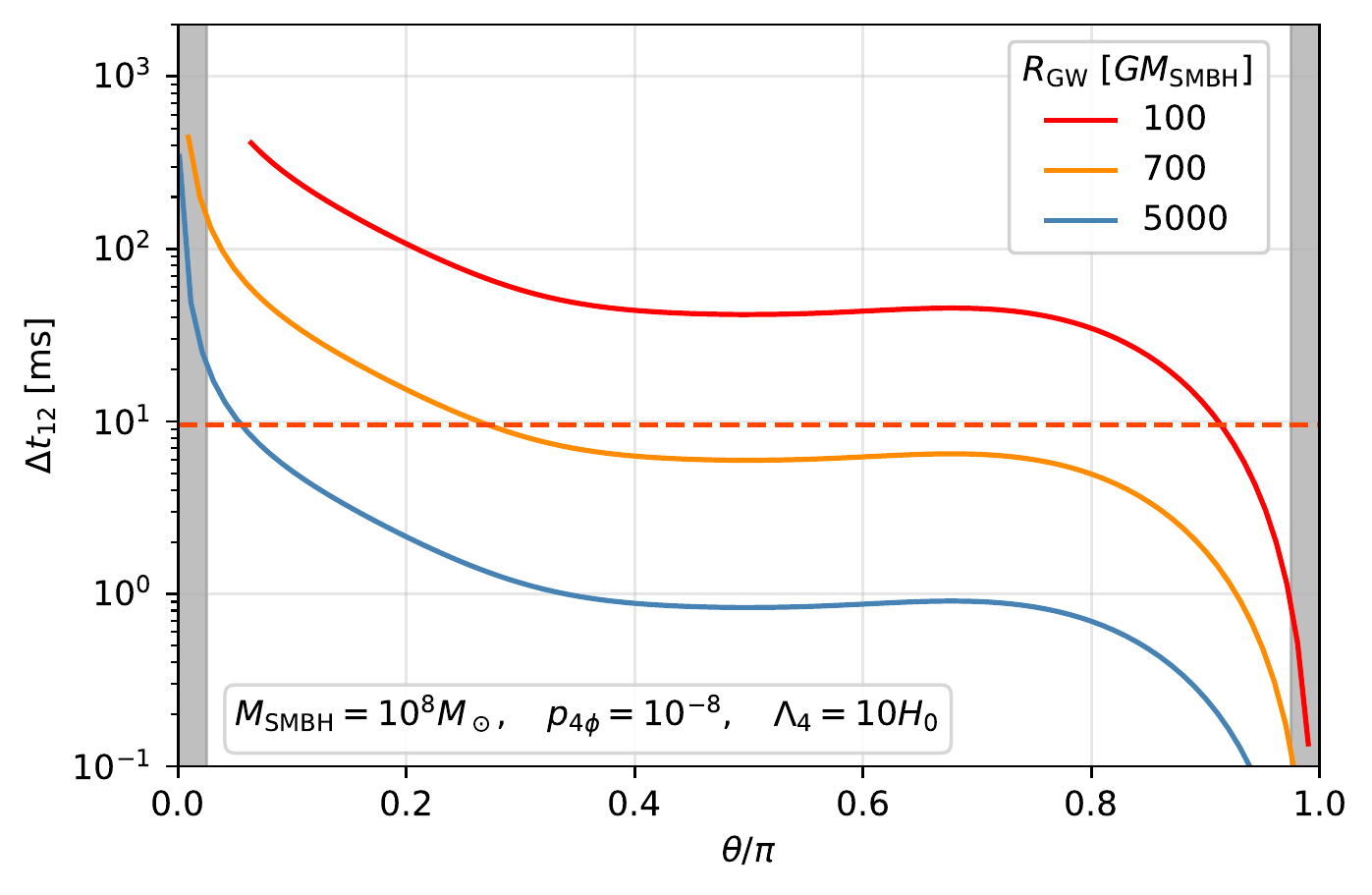}
    \caption{Birefringent time delay for a source near a SMBH as a function of the angle of the observer, relative to the SMBH. Each line corresponds to a different source distance, for model parameters compatible with GW170817  (see. Eq. ~\ref{eq:horndeski_170817}).
    The horizontal line corresponds to the lower bound on $\Delta t_{12} = 9.5 $ ms  from the analysis of GW190521. The region between the shaded areas encompasses 95\% probability for a random observer. The lowest $\theta$ represents trajectories passing at 10 Schwarzschild radii of the SMBH.
    }
    \label{fig:dt_agn_horndeski}
\end{figure}

Following Ref.~\cite{Graham:2020gwr}, we will assume an AGN binary scenario where the mass of the SMBH is $M_{\rm SMBH}\sim 10^8M_\odot$ and the source is located in a migration trap at $r\sim 700 GM_{\rm SMBH}$. Then, using the framework of Ref.~\cite{Ezquiaga:2020dao} allows us to compute the time delay as a function of the angle between the observer and the source, relative to the SMBH. The results are shown in Fig.~\ref{fig:dt_agn_horndeski} for $p_{4\phi}=10^{-8}, \Lambda_4 = 10 H_0$, compatible GW170817  \eqref{eq:horndeski_170817}, and different distances to the SMBH (the dependence on $M_{\rm SMBH}$ is less pronounced, see below). The birefringent time delay becomes very large as $\theta\to 0$\footnote{Our calculation relies on small deviations from a straight trajectory. This assumption breaks down for small angles, where one needs to consider the geodesics of the SMBH space-time instead. However, our results are conservative since actual trajectories will bend toward the SMBH, thus increasing $\Delta t_{12}$ relative to the straight propagation.}. Ultimately,  the maximum time delay is limited by the existence of the horizon, $\theta_s \approx 2GM/r$. 
The birefringence also vanishes as $\theta\to\pi$ because of geometric cancellations in spherical symmetry.

We will translate these predictions into theory parameters and include the comparison to GW190521. We will take the values of $M_{\rm SMBH}$ and the source radius fixed, and consider the credible intervals as being determined by the angle $\theta$, cf. Eq. \eqref{eq:known_lens} and Appendix~\ref{sec:beyond_poisson_stats}. 
As we do not know the emission angle, we will assume a flat prior on the sphere $P(\theta)=\sin(\theta)$, and take the upper/lower 95\% c.l. values based on $P(\theta)$ (excluding the shaded regions in Fig.~\ref{fig:dt_agn_horndeski}). 
Limits on the theory parameters can be derived by noting that $\Delta t_{12}(\theta)\propto p_{4\phi}^{4/3}  \Lambda_4^{-2/3}  M_{\rm SMBH}^{1/3}$ including different assumptions about the SMBH mass. Note that $M_{\rm SMBH}$ enters with a different scaling than the lens mass in Sec.~\ref{sec:theory_connection}, due to the source being at a fixed distance from the SMBH and within its Vainsthein radius, rather than randomly located. 

The implications of GW190521 for the example theory \eqref{eq:lagrangian_example_horndeski} are shown in Fig.~\ref{fig:horndeski_constrained}. The orange regions are excluded if we assume the AGN scenario as discussed above. The lower region excludes the $\GR$ limit $p_{4\phi}\to 0, \Lambda_4\to \infty$ and relies on trusting the measured birefringence $\Delta t_{12}\gtrsim 9.5$ ms to be due to new gravitational physics. Even if the result is interpreted as noise (e.g. Fig.~\ref{fig:190521_bg}), assuming the AGN scenario leads to exclusion of the upper orange region (assuming sensitivity to $\Delta t_{12}\lesssim 9.5$ ms). Because of the different scaling with the theory parameters, the detection of an AGN binary becomes even more constraining than GW170817 for high $\Lambda_4$. 
The beyond $\GR$ interpretation can be further probed not only by AGN events but by high-redshift multi-messenger observations. In this case, the time delay between GWs and EM counterparts scales as $\approx 1s \left(10^8 p_{4\phi}\frac{H_0}{\Lambda_4}\right)^2\frac{D}{40\rm Mpc}$ and can be probed by distant neutron-star mergers.

\section{Summary and Outlook}\label{sec:conclusion}
In this paper, we explored $\MG$  as a test of $\GR$ using observations of GWs. 
$\MG$ produces a difference in the arrival times of the GW polarisations in signals from the binary mergers, predicted by some alternatives to the $\GR$. Using the Bayesian model selection framework, not only we can identify the signatures of birefringence, but also measure the time delay between the arrival of both polarisations ($\DT$). We show that this difference can be measured with high accuracy, of order few milliseconds with existing events and is likely to improve in the future following detector upgrades. 

Using the latest GW catalog, GWTC-3, we find no strong evidence for the observation of the birefringence, with the highest $ \log \BF=3.21$ for the heaviest binary black holes so far, GW190521. 
However, after simulating similar events under different noise realizations,
we determine that there is a 
false alarm probability of 48\%.  
This event has been associated with an AGN flare, possibly indicating that the merger occurred near a SMBH. 
This AGN scenario is especially favorable for the observation of $\MG$ since the SMBH would 
act as a strong source of $\MG$. However, the AGN flare-GW association has been disputed, see e.g. 
\cite{Palmese:2021wcv}.  
Moreover, the loudness and shortness of this event makes it susceptible to different astrophysical and fundamental physics interpretations. It has also been found to 
be violating many tests of $\GR$  and mimicking many exotic scenarios of compact binary such as head-on-collision of a boson star \cite{CalderonBustillo:2020fyi} or left-right (L-R), frequency-dependent  birefringence \cite{Wang:2021gqm}. The latter effect is related to our flavour of $\MG$, with two important differences: first, L-R birefringence is defined in the basis of circularly polarized waves (left vs right, rather than $+$ vs $\times$), and second, it depends on the GW frequency. Both features also appear in the Gravitational spin Hall effect in \GR, although the L-R time-delay is very suppressed \cite{Andersson:2020gsj,Oancea:2022szu}. 

Of the 43 analyzed events, we find that the tightest bounds on the time delay between the two polarisations is $\DT \sim 0.51  $ ms at 90\% credible intervals coming from the 
GW$200311\_115853$ merger event, while the median is $\DT \simeq 80 $ ms. From the non-observation of $\MG$, we constrained the lensing optical depths in a phenomenological parameterization in which the lensing cross-section is proportional to the Einstein radius or a fixed physical radius with a power law scaling in the halo mass.

Our constraints can be translated to gravitational theories that predict $\MG$. As an example, we presented novel constraints on a Horndeski scalar-tensor theory featuring a new dynamical field and two free parameters. The theory is stringently constrained by the speed of GWs on the homogeneous FRW background following GW170817. Nevertheless, the lack of observed $\MG$ places stringent bounds, which can be orders of magnitude better than Solar System tests and in some limits as tight as the GW speed bound. 
As a proof of principle of $\MG$ due to a known inhomogeneity, we interpret GW190521 as an AGN binary (assuming that the signal originated in close proximity to a SMBH \cite{Graham:2020gwr}) in terms of our example theory. Then, the large curvature is able to generate detectable $\MG$ even when deviations from $\GR$ are minute. 
Our $|\DT|\gtrsim 9.5$ ms results would then exclude \GR, placing a minimum value of the theory parameters. When interpreting this result as a fluctuation and $\GR$ to be correct, the AGN hypothesis is still able to produce very stringent bounds, that can even overcome those of the GW speed on FRW.

In future, the methods we developed here can be useful for studying new classes of events. Of particular interest will be signals where the merger is either near an SMBH or is known to have a lensed counterpart due to strong lensing. In such cases, the information about the lens may improve the constraints substantially, along the lines of the AGN-scenario we discussed. The increase in detection rate and a growing chance of strongly lensed GW identification makes $\MG$ test also relevant for future runs of LVK detectors and upcoming GW detectors such as  Einstein Telescope, Cosmic Explorer and LISA \cite{Kalogera:2021bya, Sathyaprakash:2012jk, Ding:2015uha, mesut_LISA}. Lastly, the addition of ground-based detectors such as LIGO-India and KAGRA can allow us to measure extra linear combinations of the GW polarisations and construct a null-stream~\cite{Chatziioannou:2012rf} to extract each of the polarisations individually. The extracted polarisations can then be used to test their consistency with $\GR$ or other theories of gravity directly.

Strongly lensed GW signals may allow us to measure additional linear combinations of the same GW polarisations and hence improve various tests of $\GR$ ~\cite{Goyal_2021}, including the one proposed here. Ultimately, developing $\MG$ predictions for other alternative theories and generalizing the model-independent parameterizations presented here will allow our results to further test the landscape of theories beyond $\GR$.

\section{Acknowledgements}

We are grateful to A. K. Mehta, P. Ajith, G. Brando, S. Savastano, G. Tambalo, Y. Wang, H. Villarrubia-Rojo and J. Tasson for fruitful discussions. SG and AV are supported by the Department of Atomic Energy, Government of India, under Project No. RTI4001. AV is also supported by a Fulbright Program grant under the Fulbright-Nehru Doctoral Research Fellowship, sponsored by the Bureau of Educational and Cultural Affairs of the United States Department of State and administered by the Institute of International Education and the United States-India Educational Foundation. 
J.M.E. is supported by the European Union’s Horizon 2020 research and innovation program under the Marie Sklodowska-Curie grant agreement No. 847523 INTERACTIONS, by VILLUM FONDEN (grant no. 37766), by the Danish Research Foundation, and under the European
Union’s H2020 ERC Advanced Grant “Black holes: gravitational engines of discovery” grant agreement no. Gravitas–101052587.  
The numerical calculations reported in the paper are performed on the Alice computing cluster at ICTS-TIFR, with the aid of \textsc{LALSuite}~\cite{lalsuite}, \textsc{Bilby}~\cite{bilby}, \textsc{PyCBC}~\cite{pycbc} and \textsc{Colossus} \cite{Diemer:2017bwl} software packages. This material is based upon work supported by NSF's LIGO Laboratory which is a major facility fully funded by the National Science Foundation.

\appendix
\section{Injection Parameters}
\label{sec:app_inj}
Here we list the injection parameters for the mismatch and the parameter estimation studies. Note that the luminosity distances are scaled as per the SNRs and hence are not mentioned in the table below.
\begin{table}[h]
\begin{tabular}{llllllllll}
$m_1$ & $m_2$ & $\delta$ & $\alpha$ & $\iota$ & $\chi_1$ & $\chi_2$  & $\psi$ & $\phi_c$ & $t_c$ \\\hline
 $38.3$ & $33.19$  & $-1.2$ & $2.3$ & $2.9$ & $0.3$ & $0.27$  & $1.6$ & $1.9$ & $1126259462.414$ \\
$24.4$ & $2.7$  & $-0.4$ & $0.2$ & $0.5$ & $0.06$ & $0.46$ & $1.5$ & $4.4$ & $1249852256.99$    \\ \hline
\end{tabular}
\caption{GW150914-like (top) and GW190814-like (bottom) CBC parameters used during mismatch calculations in Sec.~\ref{sec:mismatch} and PE injection studies in Sec.~\ref{sec:results}.  }
\label{tab:injpars}
\end{table}

\begin{figure*}[htb]
\begin{minipage}[c]{0.6\linewidth}
\centering
\includegraphics[width =  \textwidth]{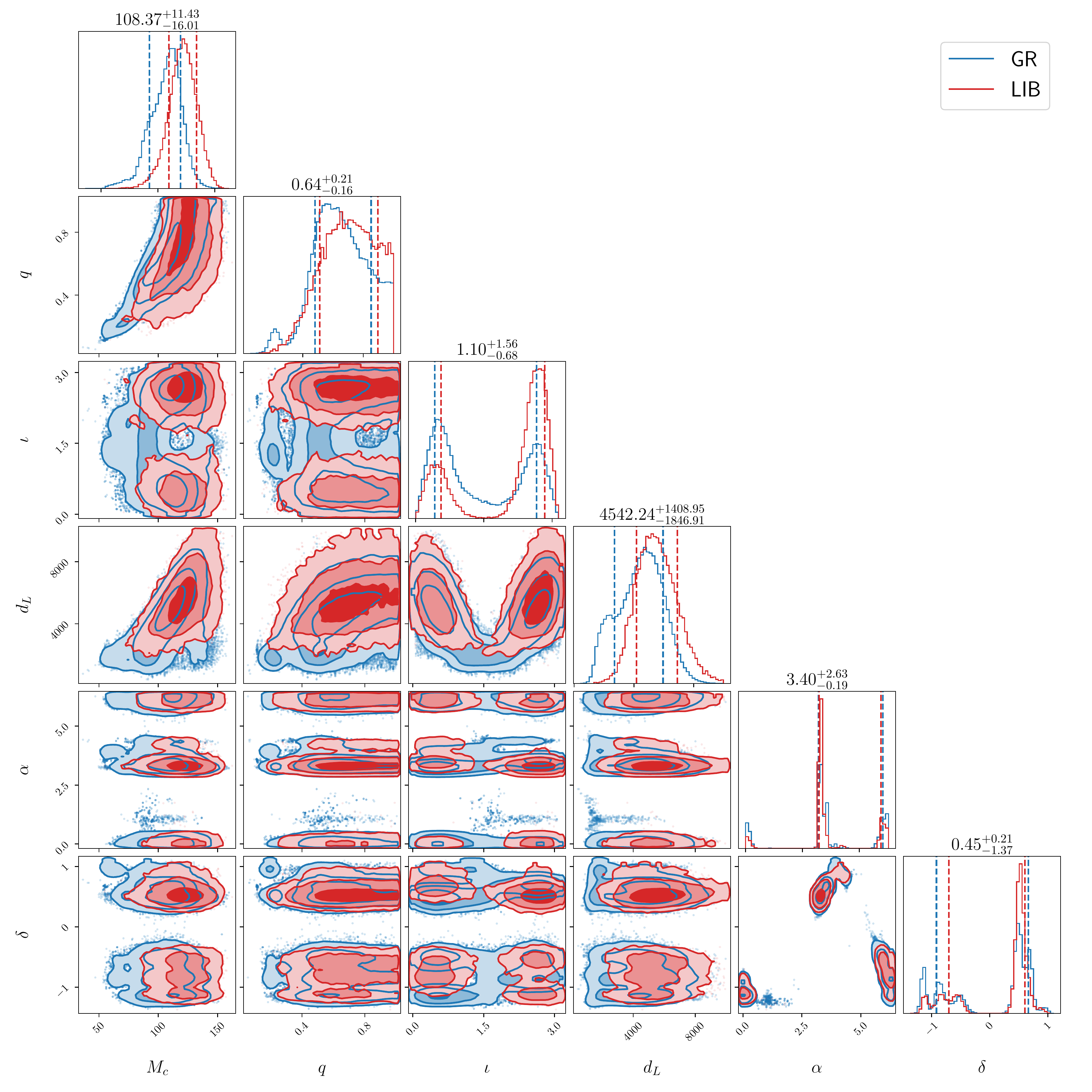} 
\end{minipage} \hfill
\begin{minipage}[c]{0.38\linewidth}
\includegraphics[trim=  0. 0 1.2cm 1.2cm ,width =  \textwidth]{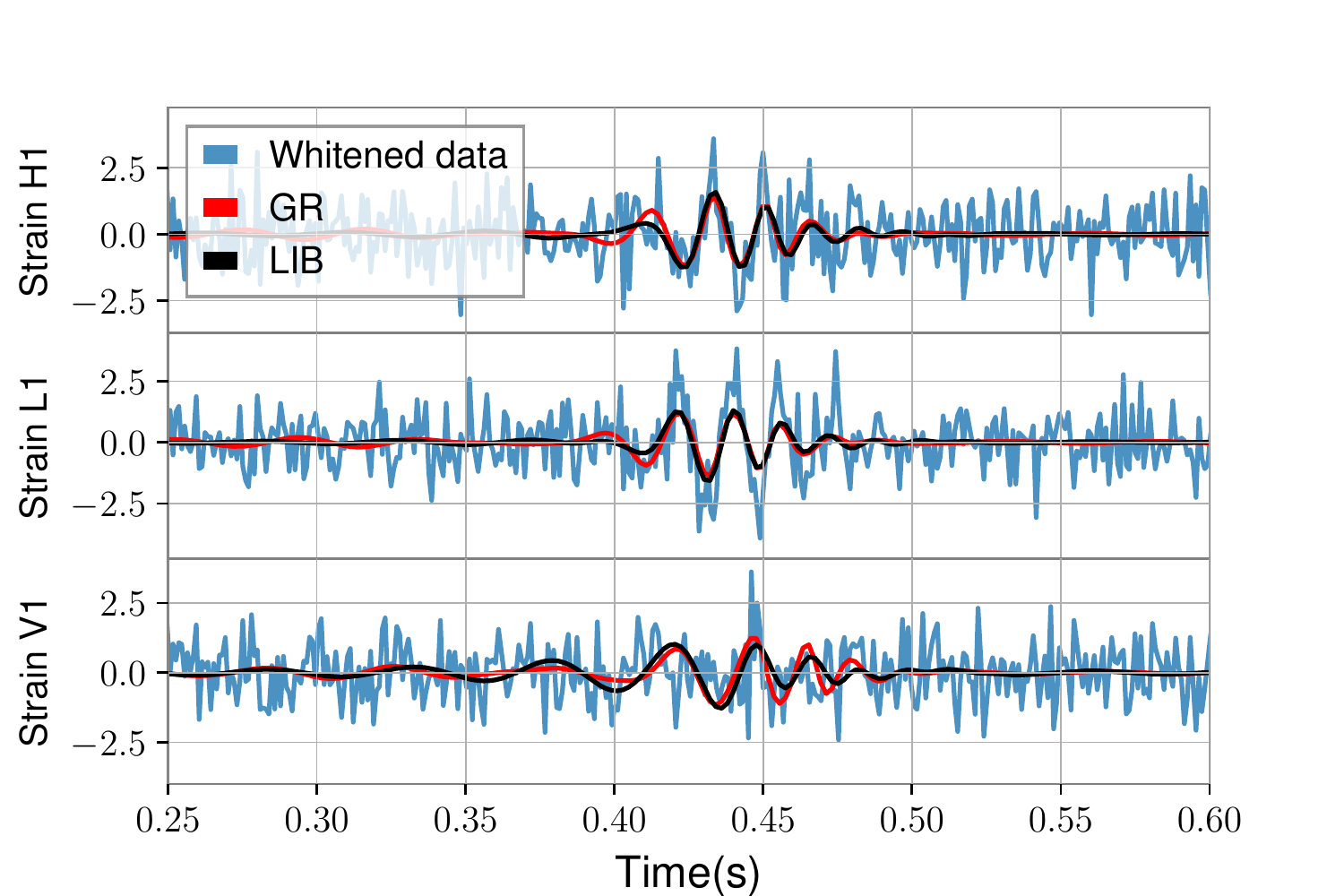}\\
\includegraphics[width =  \textwidth]{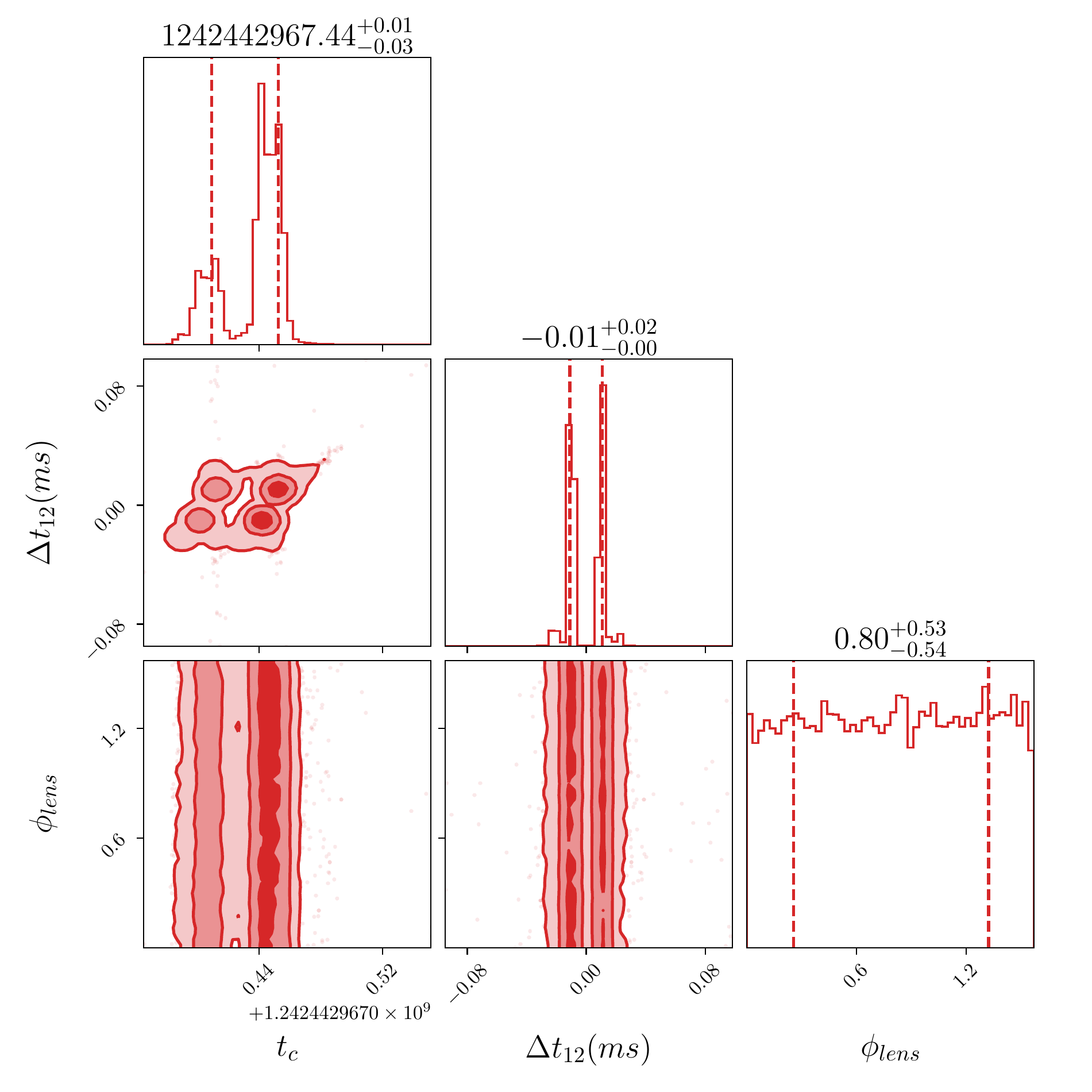}
\end{minipage}
\caption{$GW190521$($\log \BF = 3.21$)  $\GR$ v/s $\MG$ posteriors. $MaP$ (maximum a posteriori) waveforms  under $\GR$ and $\MG$ ($\DT = 9.51 \mathrm{ms}, \PL = 0.06 \mathrm{rad}$) hypothesis with the whitened strain as observed at the LIGO-Virgo detectors.  }
\label{fig:GW190521}

\end{figure*}

\section{Beyond-Poisson statistics} \label{sec:beyond_poisson_stats}

The independent lens assumption fails to capture two circumstances that potentially enhance the detection of birefringence: the source environment and lensing by known objects. This situation is qualitatively different from strong lensing probabilities, which are weighted by the Einstein radius, which vanishes when $D_L\to D_S$ (near the source) or $D_L\to 0$ (near the detector). In contrast, birefringence probabilities do not suffer such suppression and can be sizeable for objects near the source or the observer. Our optical-depth framework (Sec.~\ref{sec:cross_section_pheno}) does not consider this possibility.

Source environment may play a role for $\MG$, as GW sources will generally be located in regions denser than the cosmic average. In this case, the host galaxy (or objects within it) would have a much larger density compared to the cosmological average used in, e.g. Eq. \eqref{eq:tau_phys_halo}. In addition, the projected cross-section $\propto 1/D^2$ will be larger for nearby objects, thus enhancing the probabilities. Given a distribution of GW sources near an object, the posterior on the theory parameters $\vec p$ can be obtained as
\begin{equation}\label{eq:known_lens}
    P(\vec p) = \int dr d\theta P_{\rm s}(r)\sin(\theta) P\left(\Delta t_{12}(r,\theta,\vec p)\right)\,.
\end{equation}
Here we have assumed a symmetric $r$-dependent distribution. The $\theta$ dependence corresponds to a uniform prior on the sphere. This simple dependence could be used to model the effect of the source's galaxy or nearby objects. 

An extreme case of environmental enhancement is given by a binary merging in an AGN near a supermassive black hole (SMBH), as discussed in Sec.~\ref{sec:theory_190521}, taking the multi-messenger scenario of GW190521 and its implications for the example Horndeski theory.
Estimates for the rate of such events are uncertain. Nonetheless, in some cases it might be possible to associate an event with a SMBH thanks to an EM counterpart \cite{Graham:2020gwr}, multiple images due to strong lensing \cite{OLeary:2008myb,Kocsis:2011jy,DOrazio:2019fbq} or strong-field propagation effects \cite{Oancea:2022szu}. 

Another potential to improve the quoted result is by correlating GW arrival direction with known lenses. 
relevant in cases where the Milky Way (or perhaps even the Sun) may imprint an observable birefringence. Adding information on the GW direction, relative to known objects, will allow better constraints on those scenarios more effectively than assuming randomly located lenses. For instance, if stellar-scale lenses are relevant in a given theory and the cross-section scales as the physical radius (allowing nearby lenses to contribute), sources behind the milky way can probe a much larger effective cross-section than given by Eq.~\ref{eq:cross_section_integrals}.

Finally, any confident detection of a lensed GW can be used to refine constraints within a given model. This would follow either through the identification of several GW detections as images of the same underlying source or through  waveform distortions (millilensing). Both cases allow information about the lens mass and impact parameter to be recovered, at least when assuming a lens model \cite{Takahashi:2003ix, mesut_LISA,Tambalo:2022wlm}. That information can then place constraints within a specific theory of gravity. \\

\section{GW190521 posteriors under $\MG$ and $\GR$}
\label{sec:pe_190521}

In Fig.~\ref{fig:GW190521} we show the posteriors of the GW190521 event which has the highest log Bayes factor ($\ln \BF = 3.21$) from the PE runs of $\MG$ and $\GR$ hypothesis. The two posteriors are consistent with each other with  $\MG$ favouring a slightly higher luminosity distance ($d_L$) and chirp mass ($M_c$). 
Additionally, posteriors under $\MG$ are marginally narrower as compared to $\GR$, which might be a reason for its $ \ln \BF > 0 $. 
It is worth noticing that $\DT$ is degenerate with $t_c$, which is itself poorly measured due to low SNR in Virgo. We also plot the waveforms using maximum a posteriori (MaP) parameters  along with the whitened time series data \cite{Abbott:2020tfl} as observed in the Hanford (H1), Livingston (L1) and Virgo (V1) detectors. It's easy to see that the signal duration is small and the two MaP waveforms are not very different from each other except for the tiny modulations in the $\MG$ one. It can thus be concluded that the model selection favours the $\MG$ hypothesis because it is fitting better the random noise at the detectors during the event GW190521.   

\bibliography{gw_birefringence.bib}

\end{document}